\documentclass[12pt]{amsart}

\usepackage{epsfig}

\usepackage{amssymb,amsxtra}

\numberwithin{equation}{section}

\newcommand{\mQME}{modified quantum master equation}

\newcommand{\fullref}[1]{\ref{#1} on page~\pageref{#1}}

\newcommand{\ndash}{\nobreakdash-\hspace{0pt}}
\newcommand{\Ndash}{\nobreakdash--}
\newcommand{\tspace}[1]{\tange\! #1}
\DeclareMathOperator{\tange}{T}

\newcommand{\norm}[1]{\left\lvert\!\left\lvert#1\right\rvert\!\right\rvert}

\newcommand{\ii}{{\mathrm{i}}}
\newcommand{\dd}{{\mathrm{d}}} 
\newcommand{\ee}{{\mathrm{e}}}

\newcommand{\Omad}[1]{\Omega^{#1}(M,\mathrm{ad} P)}
\newcommand{\Omads}[1]{\Omega^{#1}(M,\mathrm{ad}^*P)}

\newcommand{\Omadf}[1]{\Omega^{#1}(N, \mathrm{ad}f^*P)}
\newcommand{\Omadsf}[1]{\Omega^{#1}(N,\mathrm{ad}^*f^*P)}

\DeclareMathOperator{\diverg}{div}
\DeclareMathOperator{\lk}{lk}

\newcommand{\mapping}[5]{{#1}\colon
  \begin{array}[t]{ccc}
{#2} &\to & {#3}\\
{#4} &\mapsto & {#5}
  \end{array}}

\newcommand{\Diff}{\mathit{Diff}}

\newcommand{\ev}{\mathrm{ev}}

\DeclareMathOperator{\tr}{Tr}

\DeclareMathOperator{\gh}{gh}

\newcommand{\salpha}{\boldsymbol\alpha}
\newcommand{\sbeta}{\boldsymbol\beta}
\newcommand{\sxi}{\boldsymbol\xi}

\newcommand{\sSigma}{\boldsymbol\Sigma}

\DeclareMathOperator{\ad}{ad}

\DeclareMathOperator{\Ad}{Ad}

\newcommand{\Lg}{\mathfrak{g}}

\newcommand{\sbv}[2]{\left(\!\left({\,{#1}\,;\,{#2}\,}\right)\!\right)}
\newcommand{\lb}[2]{[\![#1\,;#2]\!]}

\newtheorem{Thm}{Theorem}[section]
\newtheorem{Prop}[Thm]{Proposition}
\newtheorem{Lem}[Thm]{Lemma}

\theoremstyle{remark}
\newtheorem{Rem}[Thm]{Remark}
\newtheorem*{Ack}{Acknowledgment}

\theoremstyle{definition}

\newtheorem{Ass}{Assumption}
\newtheorem*{Ass*}{Assumption}

\newcommand{\braket}[2]{\left\langle{\,{#1}\,,\,{#2}\,}\right\rangle}
\newcommand{\dbraket}[2]{\left\langle\!\left\langle
{\,{#1}\ ;\,{#2}\,}\right\rangle\!\right\rangle}
\newcommand{\Lie}[2]{{\left[{\,{#1}\,,\,{#2}\,}\right]}}

\newcommand{\vev}[1]{{\left\langle\;{#1}\;\right\rangle}}

\newcommand{\bbR}{{\mathbb{R}}}

\newcommand{\de}{\partial}

\newcommand{\calA}{\mathcal{A}}
\newcommand{\calB}{\mathcal{B}}
\newcommand{\calD}{\mathcal{D}}

\newcommand{\calN}{\mathcal{N}}

\newcommand{\calI}{\mathcal{I}}
\newcommand{\calL}{\mathcal{L}}

\newcommand{\calG}{\mathcal{G}}
\newcommand{\calO}{\mathcal{O}}

\newcommand{\sfA}{{\mathsf{A}}}
\newcommand{\sfB}{{\mathsf{B}}}
\newcommand{\sfF}{{\mathsf{F}}}
\newcommand{\sfa}{{\mathsf{a}}}

\newcommand{\sfO}{{\mathsf{O}}}
\newcommand{\sfS}{{\mathsf{S}}}

\newcommand{\sfU}{{\mathsf{U}}}

\newcommand{\frg}{{\mathfrak{g}}}

\DeclareMathOperator{\bbr}{\mathbb{R}}

\DeclareMathOperator{\Imb}{Imb}

\newcommand{\sOBV}{{\boldsymbol{\Omega}}}
\newcommand{\sfdelta}{{\boldsymbol{\delta}}}
\newcommand{\sfDelta}{{\boldsymbol{\Delta}}}

\newcommand{\Imbsig}{\Imb_\sigma}

\newcommand\qq{\rm}
\newcommand\cmp[1]{{\qq Commun.\ Math.\ Phys.\ \bf #1}}
\newcommand\jmp[1]{{\qq J.\ Math.\ Phys.\ \bf #1}}
\newcommand\pl[1]{{\qq Phys.\ Lett.\ \bf #1}}
\newcommand\np[1]{{\qq Nucl.\ Phys.\ \bf #1}}

\newcommand\lmp[1]{{\qq Lett.\ Math.\ Phys.\ \bf #1}}

\begin{document} 

\title[Wilson surfaces\dots]{Wilson surfaces and 
higher dimensional knot invariants}

\author[A.~S.~Cattaneo]{Alberto S.~Cattaneo}
\address{Institut f\"ur Mathematik, Universit\"at Z\"urich--Irchel,  
Winterthurerstrasse 190, CH-8057 Z\"urich, Switzerland}  
\email{asc@math.unizh.ch}

\author[C.~A.~Rossi]{Carlo A.~Rossi}
\address{D-MATH, ETH-Zentrum, CH-8092 Z\"urich, Switzerland}
\email{crossi@math.ethz.ch}

\thanks{A.~S.~C. acknowledges partial support of SNF Grant No.~20-63821.00}

\begin{abstract}
An observable for nonabelian, higher-dimensional forms is introduced, its properties
are discussed and its expectation value in $BF$ theory is described.
This is shown to produce
potential and genuine invariants of higher-dimensional knots.
\end{abstract}

\maketitle

\section{Introduction}\label{intro}
Wilson loops play a very important role in gauge theories. They appear as 
natural observables, e.g., in Yang--Mills and in Chern--Simons theory; in the
latter, their expectation values lead to invariants for (framed) knots \cite{Wit}.
A generalization of Wilson loops in the case where the connection is replaced by a form
$B$ of higher degree and the loop by a higher-dimensional submanifold is then natural and
might have applications to the theories of D-branes, gerbes 
and---as we discuss in this paper---invariants of imbeddings.

In the abelian case, one assumes $B$ to be an ordinary $n$\ndash form on an 
$m$\ndash dimensional manifold $M$. The generalization of abelian gauge symmetries is in
this case given by transformations of the form $B\mapsto B+\dd\sigma$, 
$\sigma\in\Omega^{n-1}(M)$. 
The obvious generalization of a Wilson loop has then the form 
\begin{equation}\label{abeobs}
\calO(B,f,\lambda)=\ee^{\frac\ii\hbar \lambda\int_N f^*B},
\end{equation}
where $\lambda$ is a coupling constant, $N$ is an $n$\ndash dimensional manifold and
$f$ is a map $N\to M$.

As an example of a theory where this observable is interesting,  one has
the so-called abelian
$BF$ theory \cite{SchwBF} which is defined by the action functional 
\[
S(A,B)=\int_M B\,\dd A,\qquad
B\in\Omega^n(M),\ A\in\Omega^{m-n-1}(M).
\]
The expectation value of the product $\calO(B,f,\lambda)\calO(A,g,\Tilde\lambda)$ 
(with $f\colon N\to M$, $g\colon \Tilde N\to M$, $\dim N=n$, $\dim \Tilde N=m-n-1$)
is then an interesting topological invariant which in the case $M=\bbR^n$ turns out to be
a function of the linking number of the images of $f$ and $g$ (assuming that they do not
intersect).

A nonabelian generalization seems to require necessarily that along with $B$ one has
an ordinary connection $A$ on some principal bundle $P\to M$. 
The field $B$ is then assumed to be a tensorial $n$\ndash  form on $P$.

If the map $f(N)$ describes an $(n-1)$\ndash family
of imbedded loops (viz., $N=S^1\times X$ and $f(\bullet,x)$ is an imbedding
$S^1\hookrightarrow M$ $\forall x\in X$),
then a generalization of \eqref{abeobs} has been introduced in \cite{CCRin} in the
case $n=2$ and, more generally, in \cite{CCR,CR}. If such an observable is then considered
in the context of nonabelian $BF$ theories (which implies that one has to take $n=m-2$),
one gets cohomology classes of the Vassiliev type on the space of imbeddings of a circle
into $M$ \cite{CCR,CR,CCL}.

In the present paper we are however interested in the case where $f$ is an 
imbedding\footnote{The necessity of considering imbeddings in the nonabelian theory, instead of more
general smooth maps, arises at the quantum level (just like in the nonabelian Chern--Simons theory)
in order to avoid singularities which make the observables ill-defined.}
of $N$ into $M$. We assume throughout $n=m-2$ and we choose $B$ to be of the coadjoint type.
In particular, this will make our generalization of \eqref{abeobs},
the {\sf Wilson surface}, suitable for the
so-called canonical $BF$ theories, see Section~\ref{sec-canBF}.
Since these theories are topological, expectation values of Wilson surfaces should yield
potential invariants of imbeddings of codimension two, i.e., of
{\sf higher-dimensional knots}.

As an example,
we discuss explicitly the case when $M=\bbR^m$ and $N=\bbR^{m-2}$ and the imbeddings
are assumed to have a fixed linear behavior at infinity ({\sf long knots}). In this
case, by studying the first orders in perturbation theory, 
we recover an invariant proposed by Bott in \cite{Bott} for $m$ odd
and introduce a new invariant for $m=4$. More general invariants may be obtained at higher 
orders. (These results have appeared in \cite{Rth} to which we will recurringly refer
for more technical details.)

We believe that our Wilson surfaces may have broader applications in gauge theories.

\subsection*{Plan of the paper}
In Section~\ref{sec-canBF}, 
we recall nonabelian canonical $BF$ theories and give a very formal, but intuitively clear,
definition of Wilson surfaces, see \eqref{eq-imbaction} and \eqref{calO}. 
We discuss their formal properties and, in particular,
we clarify why we expect their expectation values to yield invariants of higher-dimensional
knots. (In this Section by invariant we mean a 
$\Diff_0(N)\times\Diff_0(M)$\ndash invariant function on the space of imbeddings
$N\hookrightarrow M$.)

In Section~\ref{sec-BVW}, we give a more precise and at the same time
more general definition of Wilson surfaces
under the simplifying assumption that we work on {\em trivial}\/ principal 
bundles. The properties of Wilson surfaces are here summarized in terms of descent 
equations \eqref{descent}, the crucial point of the whole discussion being
the \mQME\ \eqref{master}. Though we briefly recall here the fundamental facts about
the Batalin--Vilkovisky (BV) formalism \cite{BV}, some previous exposure to it will 
certainly be helpful.

In Section~\ref{sec-lk}, we carefully describe the perturbative definition of Wilson
surfaces---see \eqref{defU}, \eqref{deftausigma} and 
\eqref{expltausigma}---in the case $M=\bbr^m$ and $N=\bbr^{m-2}$ to which we will stick 
to the end of the paper. 

This perturbative definition of Wilson surfaces is finally 
rigorous, and in Section~\ref{sec-wilslong} we are able to prove
some of its properties, viz., the ``semiclassical'' version of the descent equation,
see \eqref{semicldescent} and Prop.~\ref{prop-tausigma}. The ``quantum'' descent
equation, on the other hand, still relies on some formal arguments.

In Section~\ref{sec-pertlong}, we discuss the perturbative expansion of the
expectation value of a Wilson surface in $BF$ theory. The main results
we obtain by considering the first three orders in perturbation theory are
a generalization of the self-linking number \eqref{Theta1},
the Bott invariant \eqref{Theta2}, and a new invariant for long 
$2$\ndash knots \eqref{Theta3}, see Prop.~\ref{barTheta3}.
(In this Section an invariant is understood as a locally
constant function on the space of imbeddings.)
We also discuss the general behavior of higher orders
as well as the expectation value \eqref{mixedvev} of the product of a Wilson loop and a
Wilson surface.
The discussions in this Section require some knowledge
on the compactification of configuration spaces relative to imbeddings 
described in \cite{BT}. We refer for more details on this part to \cite{Rth}.

Finally, in Section~\ref{sec-comm}, we discuss some possible extensions of our work.

\begin{Ack}
We thank J.~Stasheff for his very useful comments and for revising a first version
of the manuscript. We also thank R.~Longoni and D.~Indelicato for discussions
on the material presented here.
\end{Ack}

\section{Canonical $BF$ theories and Wilson surfaces}\label{sec-canBF}
We begin by fixing some notations that we will use throughout.
Let $G$ be a Lie group, $\frg$ its Lie algebra and
$P$ a $G$\ndash principal bundle over an $m$\ndash dimensional manifold
$M$. 
We will denote by $\calA$ and $\calG$ the affine space of connection
$1$\ndash forms and the group of gauge transformations, respectively.
Given a connection $A$ and a gauge transformation $g$, we will denote
by $A^g$ the transformed connection.
The next ingredients are the spaces $\Omad k$ and
$\Omads k$
of tensorial
$k$\ndash forms of the adjoint and coadjoint type respectively.
Given a connection $A$, we will denote
by $\dd_A$ the corresponding covariant derivatives on $\Omad \bullet$ and
on $\Omads \bullet$.

\subsection{Canonical $BF$ theories}\label{ssec-canBF}
Given $A\in\calA$ and $B\in\Omads{m-2}$, 
one defines the canonical $BF$ action functional by
\begin{equation}\label{S}
S(A,B):=\int_M \braket B{F_A},
\end{equation}
where $F_A$ is the curvature $2$\ndash form of $A$ and $\braket{\ }{\ }$ 
denotes the extension to forms of the adjoint and coadjoint type of the canonical pairing between $\frg$ and $\frg^*$.
The critical points of $S$ are pairs $(A,B)\in\calA\times\Omads{m-2}$
where $A$ is flat and $B$ is covariantly closed, i.e., solutions to
$F_A=0=\dd_A B$.

The $BF$ action functional 
is invariant under the action of an extension of the 
group $\calG$ of gauge transformations, viz., the semidirect product
$\widetilde\calG:=\calG\rtimes\Omads {m-3}$, 
where $\calG$ acts on the abelian group $\Omads{m-3}$ via the coadjoint 
action. A pair $(g,\sigma)\in\widetilde\calG$ acts on a pair
$(A,B)\in\calA\times\Omads{m-2}$ by
\begin{subequations}\label{sym}
\begin{align}
A &\mapsto A^g,\\
B &\mapsto B^{(g,\sigma)}=\Ad_{g^{-1}}^* B + \dd_{A^g}\sigma,
\end{align}
\end{subequations}
and it is not difficult to prove that $S(A^g,B^{(g,\sigma)})=S(A,B)$.

By definition an {\sf observable}
is a $\widetilde\calG$\ndash invariant function on $\calA\times\Omads{m-2}$.
In the quantum theory, one defines the expectation 
value\footnote{For notational simplicity, throughout the paper we assume
the functional measures to be normalized.}
of an 
observable 
by
\begin{equation}\label{vev}
\vev{\calO} = \int
\calD A\calD B\;
\ee^{\frac\ii\hbar S(A,B)}\;\calO(A,B),
\end{equation}
where the formal measure $\calD A\calD B$ is assumed to be 
$\widetilde\calG$\ndash invariant.

\subsection{Wilson surfaces}
We are now going to define an observable for $BF$ theories associated
to an imbedding $f\colon N\hookrightarrow M$, where $N$ is a fixed 
$(m-2)$\ndash dimensional manifold. The first observation is that, using $f$,
one can pull back the principal bundle $P$ to $N$; let us denote by $f^*P$ the
principal bundle over $N$ obtained this way. Given a connection one-form $A$
on $P$, we denote by $f^*A$ the induced connection one-form on $f^*P$;
moreover, given $B\in\Omads{m-2}$ we denote by $f^*B$ the induced element
of $\Omadsf{m-2}$. We then define
\begin{equation}\label{eq-imbaction}
\Sigma(\xi,\beta,A,B,f) :=
\int_{N} \braket{\xi}{\dd_{f^* A}\beta + f^* B},
\end{equation}
for $\xi\in\Omadf0$ and $\beta\in\Omadsf{m-3}$. Our observable, which we
will call {\sf Wilson surface}, 
is then
defined as the following functional integral:
\begin{equation}\label{calO}
\calO(A,B,f) := \int \calD\xi\calD\beta\;
\ee^{\frac\ii\hbar \Sigma(\xi,\beta,A,B,f)}.
\end{equation}
There are two important observations at this point:
\begin{enumerate}
\item At first sight we have a Gaussian integral where the quadratic part pairs
$\xi$ with $\beta$ but there is no linear term in $\beta$; so it
seems that one could omit the linear term in $\xi$ as well. As a consequence
$\calO$ would not depend on $B$ and would then have a rather trivial
expectation value in $BF$ theory. The point however is that 
\eqref{eq-imbaction} has in general zero modes. One has then to expand around
each zero mode and then integrate over them (with some measure
``hidden'' in the notation $\calD\xi\calD\beta$). This makes things
more interesting as we will see in the rest of the paper; in particular,
the dependency of $\calO$ on $B$ will be nontrivial.
\item The action functional \eqref{eq-imbaction} may have symmetries (depending
on $A$ and $B$) which make the quadratic part around critical points 
degenerate. So in the computation of $\calO$ the choice of some adapted
gauge fixing is understood. We defer a more precise discussion to 
the following Sections.
\end{enumerate}
We want now to show that (formally) $\calO$ is an observable.
First observe that an element $(g,\sigma)$ of the symmetry group 
$\widetilde\calG$
of canonical $BF$ theories, induces a pair $(\tilde g,\tilde \sigma)$,
where $\tilde g$ is a gauge transformation for $f^*P$ and
$\tilde\sigma=f^*\sigma\in\Omadsf{m-3}$. It is not difficult to show
that
\[
\Sigma(\xi,\beta,A^g,B^{(g,\sigma)},f)=
\Sigma(\Ad_{\tilde g}\xi,\Ad^*_{\tilde g}(\beta+\tilde\sigma),A,B,f).
\]
Thus, by making a change of variables in \eqref{calO}, we see that
$\calO$ is $\widetilde\calG$\ndash invariant if
we make the following
\begin{Ass}\label{ass1}
We assume that the measure $\calD\xi\calD\beta$ is invariant under
$i$) the action of gauge transformation on $\Omadf0\times\Omadsf{m-3}$
and $ii$) translations of $\beta$.
\end{Ass}
In the following we will see examples where these conditions are met; observe
that this will in particular imply conditions on the measure on zero modes.

\subsection{Invariance properties}
Next we want to discuss invariance of $\calO$ under the group $\Diff_0(N)$
of diffeomorphisms of $N$ connected to the identity. For $\psi\in\Diff_0(N)$,
one can now prove that\footnote{To be more precise, observe that the l.h.s.\ is now 
defined on tensorial
forms on $(f\circ\psi^{-1})^*P$ instead of $f^*P$. By $\psi^*$ we mean then
the isomorphism between $\calN(f):=\Omadf0\times\Omadsf{m-3}$ and
$\calN(f\circ\psi^{-1}):=\Omega^0(N,\ad(f\circ\psi^{-1})^*P)\times\Omega^{m-3}(N,\ad^*(f\circ\psi^{-1})^*P)$.}
\[
\Sigma(\xi,\beta,A,B,f\circ\psi^{-1})=
\Sigma(\psi^*\xi,\psi^*\beta,A,B,f).
\]
If we now further assume that the measure $\calD\xi\calD\beta$ is
invariant\footnote{More precisely, we assume that the measure
$\calD\tilde\xi\calD\tilde\beta$ 
on $\calN(f\circ\psi^{-1})$ is equal to the pullback
of the measure  $\calD\xi\calD\beta$ by $\psi^*$ whenever
$\tilde\xi=\psi^*\xi$ and $\tilde\beta=\psi^*\beta$.} under 
$\psi^*$, we obtain that
\[
\calO(A,B,f\circ\psi^{-1})=
\calO(A,B,f).
\]

Finally, we want to prove that 
$\vev{\calO}$ is also $\Diff_0(M)$\ndash invariant. For $\phi\in\Diff_0(M)$,
the relevant identity is now\footnote{Observe that now we are moving
from $P$ to $\phi^*P$, and in the 
r.h.s.\ $\phi^*$ denotes the induced isomorphism
between $\calA(P)\times\Omads{m-2}$ and 
$\calA(\phi^*P)\times\Omega^{m-2}(M,\ad^*\phi^*P)$.} 
\[
\Sigma(\xi,\beta,A,B,\phi\circ f)=
\Sigma(\xi,\beta,\phi^*A,\phi^*B,f).
\]
After integrating out $\xi$ and $\beta$, we get then
\[
\calO(A,B,\phi\circ f)=
\calO(\phi^*A,\phi^*B,f).
\]
Observe now that the $BF$ action \eqref{S} if $\Diff_0(M)$\ndash invariant,
viz.,
\[
S(A,B)=S(\phi^*A,\phi^*B).
\]
Thus, if we assume the measure $\calD A\calD B$ to be 
$\Diff_0(M)$\ndash invariant as well,
we deduce that $\vev\calO(f)=\vev\calO(\phi\circ f)$ 
$\forall\phi\in\Diff_0(M)$.

In conclusion, whenever we can make sense of the observable $\calO$
and the expectation value \eqref{vev}
together with assumption~\ref{ass1}, we may expect
to obtain invariants of higher-dimensional knots $N\hookrightarrow M$.
A caveat is that in the perturbative evaluation of the functional integrals
some regularizations have to be included (e.g., point splitting) and this
may spoil part of the result (analogously to what happens in Chern--Simons
theory where expectation values of Wilson loops do not actually yield knot
invariants but invariants of {\em framed}\/ 
knots\footnote{Genuine knot invariants may also
be obtained by subtracting  suitable multiples of the self-linking number \cite{BT}.
We will see in subsection~\ref{ssec-ho} that a similar strategy---viz., taking linear
combination of potential invariants coming form expectation values in order to 
obtain genuine invariant---may be used in the case of long higher-dimensional knots.}).

\subsection{The abelian case}
As a simple example we discuss now the case $\frg=\bbR$. The action $\Sigma$
simplifies  to
\[
\Sigma(\xi,\beta,A,B,f) :=
\int_{N} {\xi}(\dd\beta + f^* B).
\]
The critical points are solutions to $\dd\xi_0=\dd\beta_0 + f^* B=0$. Since we want
to treat $B$ perturbatively, we expand instead around a solution to $\dd\xi_0=\dd\beta_0=0$.
For simplicity we consider only the case $\beta_0=0$.\footnote{Observe 
that the action is invariant under the transformation $\beta\mapsto\beta+\dd\tau$.
So, if $H^{m-3}(N)=\{0\}$, there is no loss of generality in taking $\beta_0=0$.}
On the other hand $\xi_0$ has to be a constant function; we will
denote by $\Xi$ its value. We get then
\[
\calO(A,B,f) = Z \int_{\Xi\in\bbR}\mu(\Xi)\;\ee^{\frac\ii\hbar\Xi\int_Nf^*B},
\]
where $\mu$ is a measure on the moduli space $\bbR$ of solutions to $\dd\xi_0=0$, and
\[
Z= \int \calD\alpha\calD\beta\;
\ee^{\frac\ii\hbar\int_{N} {\alpha}(\dd\beta + f^* B)}=
\int \calD\alpha\calD\beta\;
\ee^{\frac\ii\hbar\int_{N} {\alpha}\dd\beta},
\]
where we have denoted by $\alpha$ the perturbation of $\xi$
around $\Xi$. Observe that $Z$ is independent 
of $f$, of $A$ and of $B$.\footnote{The 
explicit computation of $Z$, taking into account the symmetries
with the BRST formalism, yields the Ray--Singer torsion of $N$, see \cite{SchwBF}.}
If we take the measure $\mu$ to be a delta function peaked at some value $\lambda$, we recover,
apart from the constant $Z$, the observable displayed in \eqref{abeobs}.

\section{BV formalism}\label{sec-BVW}
$BF$ theories present symmetries that are reducible on shell.\footnote{The infinitesimal
form of the symmetries \eqref{sym} consists of usual infinitesimal gauge symmetries and
of the addition to $B$ of the covariant derivative of an $(m-3)$\ndash form $\sigma$
of the coadjoint type. On shell, i.e.\ at the critical points of the action, the connection
has to be flat. Thus, there is a huge kernel of infinitesimal symmetries containing
in particular all $\dd_A$\ndash exact forms. Off shell the kernel is in general much smaller.
Having completely different kernels on and off shell makes the BRST formalism, even with
ghosts for ghosts, not applicable to this case.}
To deal with it,
one resorts to the Batalin--Vilkovisky (BV) formalism. 
We summarize here the results on BV for canonical $BF$ theories \cite{CR}.
First we introduce the following spaces of superfields:
\begin{align*}
\boldsymbol\calA &:= \calA \oplus \bigoplus_{\substack{i=0\\i\not=1}}^m
\Omad i [1-i],\\
\boldsymbol\calB &:= \bigoplus_{i=0}^m \Omads i [m-2-i],
\end{align*}
where the number in square brackets denotes the ghost number to be
given to each component. If we introduce the {\sf total degree} as the sum of
ghost number and form degree, we see that elements of $\calA$ have total
degree equal to one and elements of $\calB$ have total degree equal
to $m-2$. 
\begin{Rem}
In the following, whenever we refer to some super algebraic 
structure(Lie brackets, derivations,\dots), it will always be understood
that the grading is the total degree.
\end{Rem}
Observe then that the space $\boldsymbol\calA$
of superconnections is modeled on the super vector space
\[
\boldsymbol\calA_0:=
\bigoplus_{i=0}^m
\Omad i [1-i].
\]
The Lie algebra structure on $\frg$ induces a super Lie 
algebra structure on $\boldsymbol\calA_0$ whose Lie bracket will be denoted
by $\lb{\ }{\ }$.
(We refer to \cite{CR} for more details and sign 
conventions.)\footnote{It suffices here to say that (locally) the Lie bracket of
$\frg$\ndash valued forms $\alpha$ and $\beta$ is defined by
\[
\lb\alpha\beta=(-1)^{\gh\alpha\deg\beta}\,\alpha^a\beta^b\,f_{ab}^c\,R_c,
\]
where $\{R_c\}$ is a basis of $\frg$, $f_{ab}^c$ are the corresponding 
structure constants, $\gh$ denotes the ghost number and $\deg$ the form degree.}
Given $\sfA\in\boldsymbol\calA$, we define its curvature
\[
\sfF_\sfA= F_{A_0} + \dd_{A_0}\sfa + \frac12\lb\sfa\sfa,
\]
where $A_0$ is any reference connection and 
$\sfa:=\sfA-A_0\in\boldsymbol\calA_0$.
Then we define the BV action for the canonical $BF$ theory by
\[
\sfS(\sfA,\sfB)=\int_{M} \dbraket{\sfB}{\sfF_{\sfA}},
\]
where $\dbraket{\ }{\ }$ denotes the extension to forms of the adjoint and coadjoint type 
of the canonical pairing between $\frg$ and $\frg^*$ with shifted 
degree:
\[
\dbraket\alpha\beta:=(-1)^{\gh\alpha\deg\beta}\braket\alpha\beta.
\]
Integration over $M$ is assumed here to select the form component of degree $m$.
Observe that $\sfS(A,B)=S(A,B)$ as in \eqref{S}.

The
space $\boldsymbol\calA\times\boldsymbol\calB$ of superfields is isomorphic
to $\tspace^*[-1]\boldsymbol\calA$ and as such it has a canonical odd symplectic
structure whose corresponding BV bracket we will denote by $\sbv{\ }{\ }$.
It can then be shown that $\sfS$ satisfies the classical master equation
$\sbv\sfS\sfS=0$. This implies that the derivation (of total degree one)
$\boldsymbol{\delta}:=\sbv\sfS{\ }$ is a differential (the BRST differential).
It can be easily checked that
\begin{equation}\label{sfdelta}
\sfdelta\sfA = (-1)^m\,\sfF_\sfA,\qquad
\sfdelta\sfB = (-1)^m\,\dd_\sfA\sfB.
\end{equation}
As usual in the BV formalism one also introduces the BV Laplacian
$\sfDelta$. For this, one assumes a measure which induces a divergence
operator and defines
$\sfDelta F$ by $\frac12\diverg X_F$ with $X_F=\sbv F{\ }$ the Hamiltonian vector field
of $F$. In the functional integral, the measure is defined only formally.
For us, the Laplace operator will have the property that
\begin{equation}\label{DeltaAB}
\sfDelta((\sfA_k)^a(x)\,(\sfB_l)_b(y)) = \delta_{k+l,-1}\,\delta^a_b\,
\delta(x,y),
\end{equation}
where $\sfA_k$ ($\sfB_k$) denotes the component of ghost number $k$
of  $\sfA$ ($\sfB$), and we have chosen a local trivialization of
$\ad P$ ($\ad^* P$) to expand $\sfA_k$ ($\sfB_k$) on a basis of $\frg$ 
($\frg^*$).
One can then show that $\sfDelta\sfS=0$. As a consequence $\sfS$ satisfies
the quantum master equation $\sbv\sfS\sfS-2\ii\hbar\sfDelta\sfS=0$,
and the operator
\[
\sOBV:=\sfdelta-\ii\hbar\sfDelta
\]
is a coboundary operator (i.e., $\sOBV^2=0$) of total degree one.

Given a function $\sfO$ on $\tspace^*[-1]\boldsymbol\calA$, one defines
its expectation value by
\[
\vev\sfO := \int_\calL
\calD \sfA\calD\sfB\;
\ee^{\frac\ii\hbar \sfS(\sfA,\sfB)}\;\sfO(\sfA,\sfB),
\]
where $\calL$ is a Lagrangian submanifold (determined by a 
gauge fixing). The general properties of the BV formalism ensure that
\begin{enumerate}
\item the expectation value of an $\sOBV$\ndash closed function
(called a {\sf BV observable})
is invariant under deformations of $\calL$ 
(``independence of the gauge fixing''); and
\item  the expectation value of an $\sOBV$\ndash exact function
vanishes (``Ward identities'').
\end{enumerate}

\subsection{Wilson surfaces in the BV formalism}\label{ssec-WsBV}
We want now to extend the observable $\calO$ to a function $\sfO$
(of total degree zero)
on $\tspace^*[-1]\boldsymbol\calA\times\Omega^\bullet(\Imb(N,M))$ (where
$\Imb(N,M)$ denotes the space of imbeddings $N\hookrightarrow M$)
that satisfies the ``descent equations''
\begin{equation}\label{descent}
\sOBV\sfO = (-1)^m\,\dd\sfO,
\end{equation}
where $\dd$ is the de~Rham differential on $\Omega^\bullet(\Imb(N,M))$.
Observe that denoting by $\sfO_i$ the $i$\ndash form component,
the descent equation implies in particular
\begin{align*}
\sOBV\sfO_0 &= 0,\\
\sOBV\sfO_1 &=(-1)^m\,\dd\sfO_0.
\end{align*}
Thus, $\sfO_0$ will be a BV observable satisfying $\dd\vev{\sfO_0}=0$.
We expect then that (apart from regularization problems) $\vev{\sfO_0}$
should yield a higher-dimensional knot invariant.
Observe that, since $\sfO$ will be defined in terms of a gauge-fixed
functional integral, we will have to take care of the dependence of $\sfO$
under the gauge fixing. 
We will show that the variation of $\sfO$ w.r.t.\ the
gauge fixing is $(\dd+(-1)^m\sOBV)$\ndash exact. 
As a consequence,
the variation of $\sfO$ w.r.t.\ the
gauge fixing will be $\dd$\ndash exact
and hence well defined in
cohomology. In particular, we should expect that $\vev{\sfO_0}$ should
be gauge-fixing independent.

In order to define $\sfO$ properly and to show its properties we make from
now on the following simplifying
\begin{Ass}
We assume that the principal bundle $P$ is trivial. As a consequence,
from now on, elements of $\boldsymbol\calA$ ($\boldsymbol\calB$) will
be regarded as forms on $M$ taking values in $\frg$ ($\frg^*$).
\end{Ass}

Our definition of $\sfO$ requires first the introduction of superfields
on $N$. We set
\begin{align*}
\boldsymbol{\widehat\calA} &:= \bigoplus_{i=0}^{m-2}
\Omega^i(N;\frg) [-i],\\
\boldsymbol{\widehat\calB} &:= \bigoplus_{i=0}^{m-2} \Omega^i(N;\frg^*) [m-3-i].
\end{align*}
Elements of $\boldsymbol{\widehat\calA}$ have then total degree zero, while elements of
$\boldsymbol{\widehat\calB}$ have total degree $m-3$.
Again we may regard $\boldsymbol{\widehat\calA}\times\boldsymbol{\widehat\calB}$ as
$\tspace^*[-1]\boldsymbol{\widehat\calA}$, which we endow with its canonical odd symplectic
structure. We will denote by $\sbv{\ }{\ }\sphat$ the corresponding BV bracket.

We are now in a position to give a first BV generalization of 
\eqref{eq-imbaction}; viz., for $\sxi\in\boldsymbol{\widehat\calA}$
and $\sbeta\in\boldsymbol{\widehat\calB}$, we define
\[
\sSigma_0(\sxi,\sbeta,\sfA,\sfB)(f) :=
\int_{N} \dbraket{\sxi}{\dd_{f^* \sfA}\sbeta + f^* \sfB}.
\]
One can immediately verify that 
$\sSigma_0(\xi,\beta,A,B)(f)=\Sigma(\xi,\beta,A,B,f)$.

The notation used suggests that we want to consider
$\sSigma_0(\sxi,\sbeta,\sfA,\sfB)$ as a function on $\Imb(N,M)$.
More generally, we want to define a functional $\sSigma$ taking
values in forms on $\Imb(N,M)$. To do so, we first introduce the evaluation
map
\[
\ev\colon\
\begin{array}[t]{ccc}
N\times\Imb(N,M) &\to & M\\
(x,f) &\mapsto & f(x),
\end{array}
\]
and the projection $\pi\colon N\times\Imb(N,M)\to \Imb(N,M)$.
Denoting by $\pi_*$ the corresponding integration along the fiber $N$,
we define
\[
\sSigma(\sxi,\sbeta,\sfA,\sfB) :=
\pi_* \dbraket{\sxi}{\dd_{\ev^* \sfA}\sbeta + \ev^* \sfB}
\in \Omega^\bullet(\Imb(N,M)).
\]
Observe that $\sSigma$ is a sum of forms on $\Imb(N,M)$ of different ghost 
numbers with total degree equal to zero and
that $\sSigma_0$ is the component of $\sSigma$ of 
form degree zero (or, equivalently, of ghost number zero).
Now, by using \eqref{sfdelta} and the property $\dd\pi_*=(-1)^m\pi_*\dd$,
one can prove the identity\footnote{Observe
that, in order to compute $\sbv\sSigma\sSigma\sphat$, one has to
``integrate by parts.'' This is allowed since $\dbraket\sxi\sbeta$
does not depend on the given imbedding. As a consequence, 
$\pi_*\dbraket\sxi\sbeta$ is a constant zero-form on $\Imb(N,M)$, which
implies the useful identity
\[
0=(-1)^m\,\dd\pi_*\dbraket\sxi\sbeta=
\pi_*\dbraket{\dd_\sfA\sxi}\sbeta +
\pi_*\dbraket\sxi{\dd_\sfA\sbeta}.
\]}
\begin{equation}\label{cmaster}
\dd\sSigma = (-1)^m\,\sfdelta\sSigma + \frac12\,\sbv\sSigma\sSigma\sphat.
\end{equation}
We may also define the derivation $\boldsymbol{\widehat\sfdelta}:=\sbv{\sSigma}{\ }\sphat$ which,
by \eqref{cmaster} is not a differential; on generators it gives
\begin{equation}\label{delta'}
\boldsymbol{\widehat\sfdelta}\sxi = (-1)^m\,\dd_{\ev^*\sfA}\sxi,\qquad
\boldsymbol{\widehat\sfdelta}\sbeta = (-1)^m\,\left(\dd_{\ev^*\sfA}\sbeta+\ev^*\sfB\right).
\end{equation}
Observe that for any given family of imbeddings, one gets a vector
field on $\tspace^*[-1]\boldsymbol{\widehat\calA}$.

We now introduce a formal measure $\calD\sxi\calD\sbeta$
on this space. In terms of this measure, we define
the BV Laplacian $\boldsymbol{\widehat\Delta}$. 
We assume the formal measure to
satisfy the following generalization of 
Assumption~\fullref{ass1}:\
\begin{Ass}\label{ass2}
We assume the measure to be invariant under the vector fields
defined by \eqref{delta'}; viz., we assume
$\boldsymbol{\widehat\Delta}\sSigma=0$.
\end{Ass}
Formally we can now improve \eqref{cmaster} to the fundamental identity
of this theory which we will call the {\sf modified quantum master equation};
viz,
\begin{equation}\label{master}
\boxed{\dd\sSigma = 
(-1)^m\,\left(\sOBV\sSigma + \frac12\,\sbv\sSigma\sSigma\right)
+ 
\frac12\,\widehat{\mathrm{QME}}(\sSigma)}
\end{equation}
with
\[
\widehat{\mathrm{QME}}(\sSigma) := \sbv\sSigma\sSigma\sphat -2\ii\hbar\boldsymbol{\widehat\Delta}\sSigma.
\]
This identity is a consequence of the following formal facts:
\begin{enumerate}
\item $\sfDelta\sSigma$ vanishes 
since $\sSigma$ is at most linear in $\sfA$ and $\sfB$;
\item $\boldsymbol{\widehat\Delta}\sSigma$ vanishes by Assumption~\ref{ass2}.
\item $\sbv\sSigma\sSigma$ is proportional to a delta function
at coinciding points, 
but the coefficient is proportional to 
$\dbraket{\lb\sxi\sxi}\sbeta$ which
vanishes since $\sxi$ has total degree zero. 
\end{enumerate}
Observe finally that the \mQME\ can also be rewritten in the form
\begin{equation}\label{master'}
\left(
\dd-(-1)^m\,\sOBV +\ii\hbar\boldsymbol{\widehat\Delta}
\right) \ee^{\frac\ii\hbar\sSigma}=0.
\end{equation}

We are now in a position to define the observable $\sfO$ and to prove
its formal properties. We set
\[
\sfO_\Psi:=\int_{\calL_\Psi} \calD\sxi\calD\sbeta\;
\ee^{\frac\ii\hbar\sSigma},
\]
where $\calL_\Psi$ is the Lagrangian section determined by the
gauge-fixing fermion $\Psi$. Recall that, as in general in the BV formalism,
$\Psi$ is required to depend only on the 
fields.\footnote{As usual one has first to
enlarge the space of fields and antifields by adding enough
antighosts $\bar\sigma_i$ and Lagrange
multipliers $\lambda_i$ together with their antifields
$\bar\sigma_i^+$ and $\lambda_i^+$. One then 
extends the action functional $\sSigma$ by adding the term
$\sum_i\int_N\bar\sigma_i^+\lambda_i$. The extended action still satisfies
the {\mQME}. The gauge-fixing fermion is assumed to depend on
the fields only, i.e., on the
$\bar\sigma_i$s, the $\lambda_i$s, and
the components of nonnegative ghost number in $\sxi$ and $\sbeta$.
See, e.g., subsection~\ref{ssec-gf}.}
In this modified situation, we call {\sf good} a gauge-fixing fermion
that in addition satisfies the equation
\[
\sOBV\Psi + \sbv\Psi\sSigma = (-1)^m\,\dd\Psi.
\]
In particular, gauge-fixing fermions independent of $\sfA$, $\sfB$ 
and the imbedding are 
good. 

Now let $\Psi_t$ be a path of good gauge-fixing fermions. By the usual
manipulations in the BV formalism, the \mQME\ (\ref{master}) implies that
\[
\frac\dd{\dd t}\calO_{\Psi_t} =
\frac\ii\hbar\,
((-1)^m\sOBV-\dd)\widetilde\calO_{\Psi_t}
\]
with
\[
\widetilde\calO_{\Psi_t} =
\int_{\calL_{\Psi_t}} \calD\sxi\calD\sbeta\;
\ee^{\frac\ii\hbar\sSigma}\;
\frac\dd{\dd t}\Psi_t.
\]
As a consequence, the expectation value of $\calO_\Psi$ will be 
gauge-fixing independent modulo exact forms on $\Imb(N,M)$ as long
as we stay in the class of good gauge fixings. This understood, from now
on we will drop the label $\Psi$. 

Another consequence of the \mQME\
are the descent equations \eqref{descent}, which are immediately obtained
by integrating \eqref{master'} over the Lagrangian section $\calL_\Psi$
determined by a good gauge-fixing fermion $\Psi$.

\section{The case of long higher-dimensional knots}\label{sec-lk}
We will concentrate from now on on the case $M=\bbR^m$ and 
$N=\bbR^{m-2}$, $m>3$.
We also choose once and for all a reference linear imbedding
$\sigma\colon\bbR^{m-2}\hookrightarrow\bbR^m$ and 
we consider only those imbedding that outside a compact coincide
with $\sigma$;
we denote by $\Imbsig$ the corresponding space, whose elements
are usually called long $(m-2)$\ndash knots.

On the trivial bundle $P\simeq\bbR^m\times G$, we pick the trivial 
connection as a reference point. Thus, we may identify $\calA$
with the space of $\frg$\ndash valued $1$\ndash forms.
More generally, we think of $\boldsymbol\calA$ and
$\boldsymbol\calB$ as spaces of $\frg$\ndash\ resp.\ $\frg^*$\ndash valued
forms. Observe that the pair $(A,B)=(0,0)$ is now a critical point
of $BF$ theory. We will denote by $\sfa$ and $\sfB$ the perturbations
around the trivial critical point, but, in order to keep track that
they are ``small'', we will scale them by $\sqrt\hbar$. Observe that
we assume the fields $\sfa$ and $\sfB$ to vanish at infinity.
To simplify the following computations, we also rescale 
$\sxi\to\hbar\sxi$.
As a consequence,
the super $BF$ action functional and the super $\Sigma$ functional will now
read as follows:
\begin{subequations}\label{SSigma}
\begin{align}
\frac1\hbar\,
\sfS(\sfa,\sfB) &= \int_M \dbraket\sfB{\dd\sfa + \frac{\sqrt\hbar}2\,
\lb\sfa\sfa},\label{SSigma-S}\\
\frac1\hbar\,\sSigma(\sxi,\sbeta,\sfa,\sfB) &=
\pi_* \dbraket{\sxi}{\dd\sbeta}+
\sqrt\hbar\,
\pi_* \dbraket{\sxi}{\lb{\ev^* \sfa}\sbeta + \ev^* \sfB}.\label{SSigma-Sigma}
\end{align}
\end{subequations}

\subsection{Zero modes}
We now consider the critical points of $\sSigma$ for $\hbar=0$. 
The equations of motions are simply
$\dd\xi=\dd\beta=0$.
Using translations by exact forms (which are the symmetries for $\sSigma$ at
$\hbar=0$), a critical point can always be put in the form  $\beta=0$ and
$\xi$ a constant function, whose value we will denote by $\Xi\in\frg$. 
We have now to choose a measure $\mu$ on the space $\frg$
of zero modes. 
Then we write $\sxi=\Xi+\salpha$ with $\salpha$ assumed to vanish at infinity.
We also assume $\sbeta$ to vanish at infinity and write
\[
\sfO(\sfA,\sfB) = \int_{\Xi\in\frg}\mu(\Xi)\; \sfU(\sfA,\sfB,\Xi),
\]
with 
\begin{equation}\label{defsfU}
\sfU(\sfA,\sfB,\Xi):=\int_{\calL_\Psi} \calD\salpha\calD\sbeta\;
\ee^{\frac\ii\hbar\sSigma(\Xi+\salpha,\sbeta,\sfA,\sfB)}.
\end{equation}
In the following, we will concentrate on $\sfU(\sfA,\sfB,\bullet)$
which we will regard as an element of the completion of the symmetric
algebra of $\frg^*$.

Before starting the perturbative expansion of $\sfU$, we comment briefly
on the validity of Assumption~\fullref{ass2}. 
We assume the formal measure
$\calD\salpha\calD\sbeta$ to be induced from a given constant measure
on $\frg$. This means that $\boldsymbol{\widehat\Delta}$ will have the following
property (cf.\ with \eqref{DeltaAB} for notations):
\begin{equation}\label{Delta'}
\boldsymbol{\widehat\Delta}((\sxi_k)^a(x)\,(\sbeta_l)_b(y)) = \delta_{k+l,-1}\,\delta^a_b\,
\delta(x-y).
\end{equation}
Then,
by a computation analogous to that for canonical
$BF$ theories, one obtains in $\boldsymbol{\widehat\Delta}\sSigma$
a combination of delta functions and its 
derivatives at coinciding points (!) 
but with a vanishing coefficient.
So, formally, Assumption~\ref{ass2} is satisfied.\footnote{If 
we think in terms
of the vector fields defined by \eqref{delta'}, we should take care only
of the terms containing the covariant derivatives as the formal measure
is, as usual, assumed to be translation invariant. If the Lie algebra
$\frg$ were unimodular, then we would immediately
conclude that, formally, the measure is invariant under this generalized
gauge transformation. However, even more formally, things work in general
as the contributions of different field components cancel each other.}

\subsection{The Feynman diagrams}
\begin{figure}[h!]
\begin{center}
\resizebox{9.truecm}{!}{\includegraphics{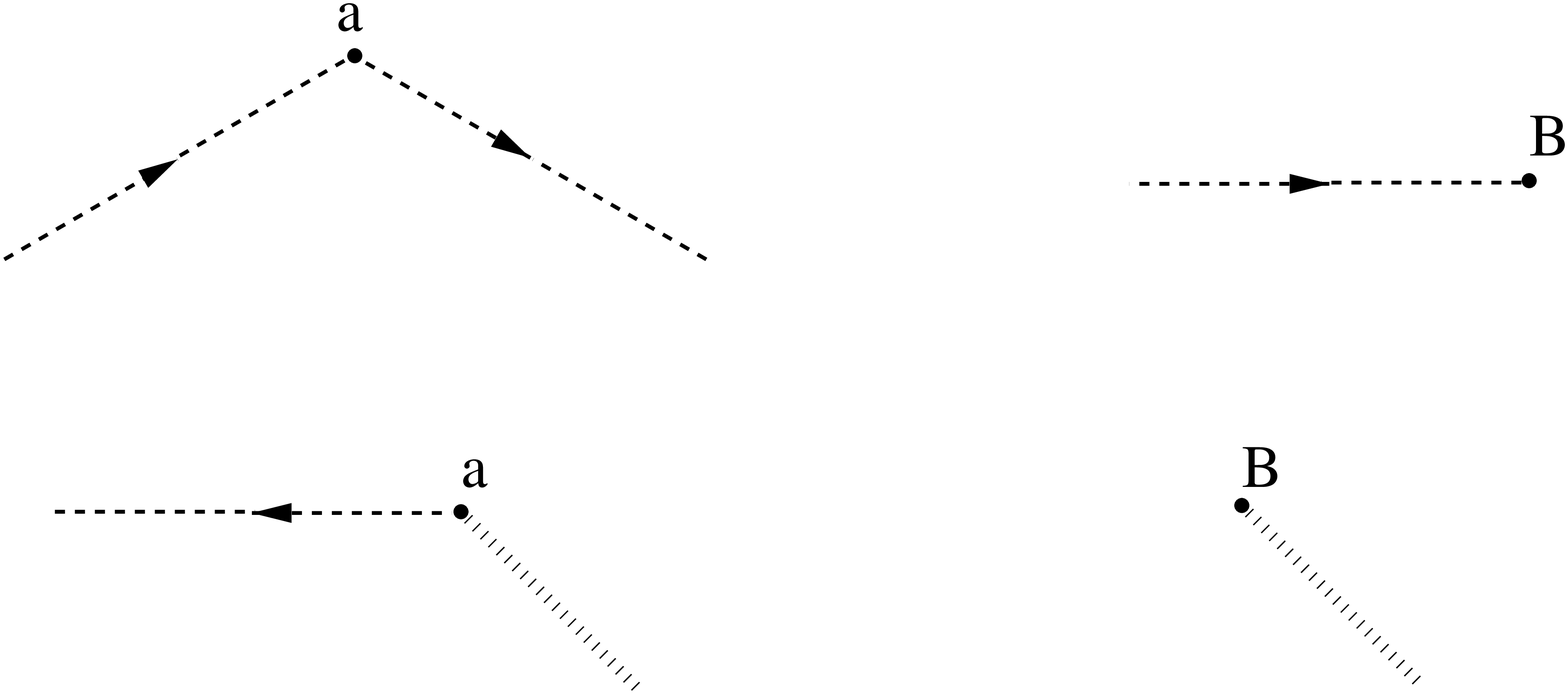}}
\caption{The four vertices coming from the second equation 
in (\ref{eq-sSigma}).}
\label{fig-vert}
\end{center}
\end{figure}

We split the action $\sSigma(\Xi+\salpha,\sbeta,\sfA,\sfB)$ into the sum of 
$\sSigma^{(0)}(\salpha,\sbeta)$ and the perturbation
$\sSigma^{(1)}_\Xi(\salpha,\sbeta,\sfA,\sfB)$:
\begin{equation}\label{eq-sSigma}
\begin{aligned}
\frac1\hbar\,\sSigma^{(0)}(\salpha,\sbeta) &=
\pi_* \dbraket{\salpha}{\dd\sbeta} =
\int_{\bbR^{m-2}}\dbraket{\salpha}{\dd\sbeta},\\
\frac1\hbar\,\sSigma^{(1)}_\Xi(\salpha,\sbeta,\sfa,\sfB) &=
\sqrt\hbar
\,\pi_* 
(\dbraket{\salpha}{\lb{\ev^* \sfa}\sbeta}+
\dbraket{\salpha}{\ev^* \sfB}+\\
&\phantom{=} +
\dbraket{\Xi}{\lb{\ev^* \sfa}\sbeta}+
\dbraket{\Xi}{\ev^* \sfB}).
\end{aligned}
\end{equation}
As a consequence, in the perturbative expansion of $\sfU$, we will have
a propagator of order $1$ in $\hbar$ (the inverse of $\dd$ with some gauge
fixing) and four vertices of order $\sqrt\hbar$.
Graphically, we will denote the propagator by a dashed line oriented
from $\sbeta$ to $\salpha$. The four vertices are then represented as in 
fig.~\ref{fig-vert}, where the black and white strip represents the zero mode $\Xi$.

Observe that with these vertices one can construct
two types of connected diagrams:
\begin{enumerate}
\item Polygons consisting only of vertices of the first type, see 
fig.~\ref{fig-poly} (observe that the $1$\ndash gon is a tadpole, so in
general it will be removed by renormalization);
\begin{figure}[h!]
\begin{center}
\resizebox{11.truecm}{!}{\includegraphics{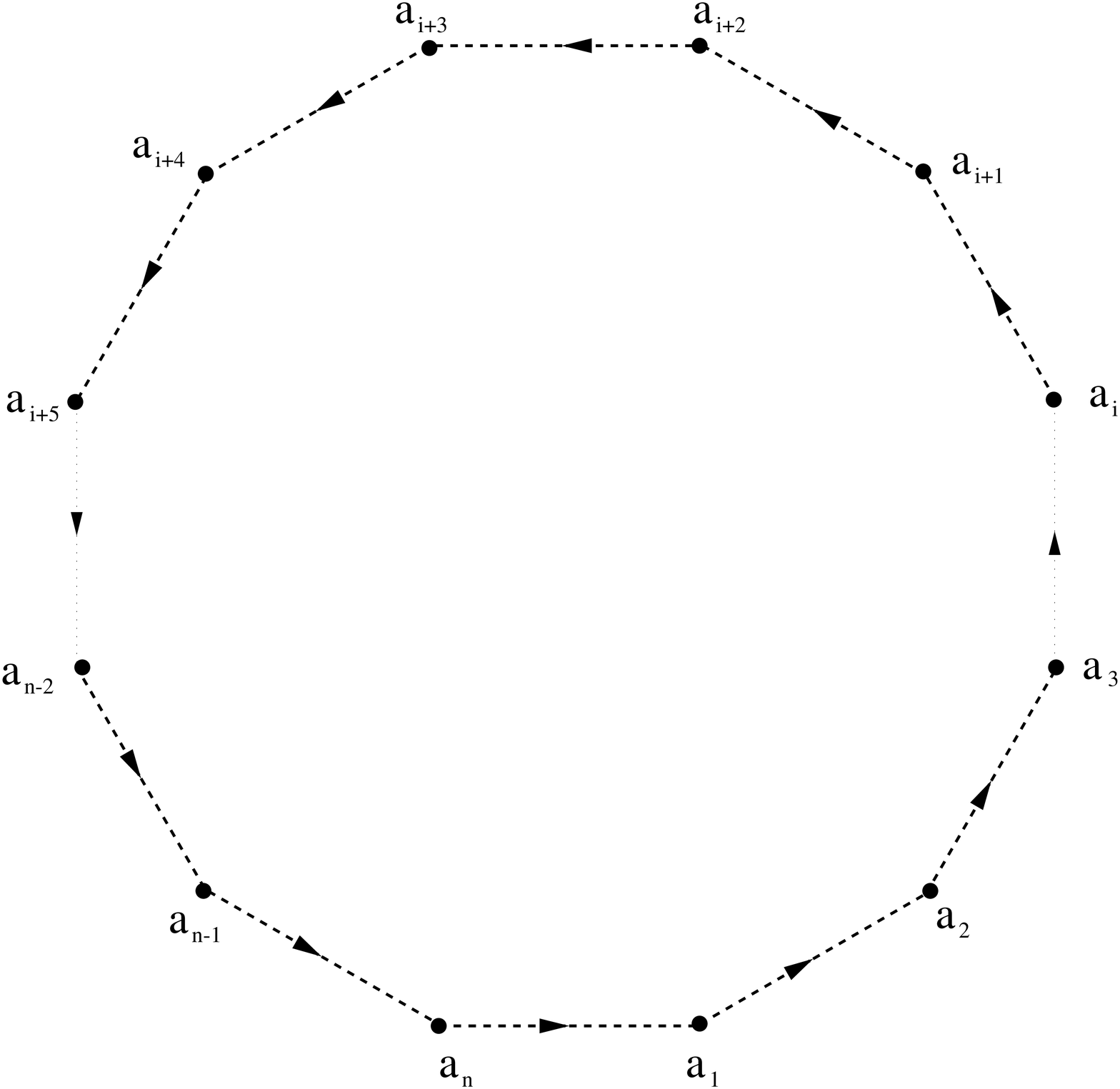}}  
\caption{The polygon $\tau_n$ with $n$ vertices.}
\label{fig-poly}
\end{center}
\end{figure}
\item ``Snakes'' with a $\sfB$\ndash field at the head and a zero mode
at the tail; there is  a very short snake consisting of a vertex of the
fourth type only; a longer snake consisting of a vertex of the second type
followed by a vertex of the third type; and a sequel of longer snakes
consisting of a vertex of the second type
followed by vertices of the first type and ending with  
a vertex of the third type. See fig.~\ref{fig-partransp}.
\begin{figure}[h!]
\begin{center}
\resizebox{12.truecm}{!}{\includegraphics{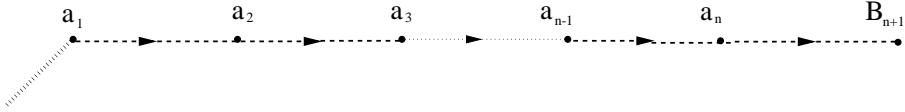}}  
\caption{The ``snake'' $\sigma_n$ with $n+1$ terms.}
\label{fig-partransp}
\end{center}
\end{figure}
\end{enumerate}
We will denote by $\tau_n$ the $n$\ndash gon and by $\sigma_n$ the
snake with $n$ vertices beside the head. Then, the combinatorial structure
of $\sfU$ is given by
\begin{equation}\label{defU}
\sfU = \ee^{\sigma+\tau},
\end{equation}
with
\begin{equation}\label{deftausigma}
\tau = \sum_{n=2}^\infty \frac{\hbar^{\frac n2}}n\, \tau_n,\qquad
\sigma = \sum_{n=0}^\infty \hbar^{\frac{n+1}2}\, \sigma_n.
\end{equation}
(The factor $n$ dividing $\tau_n$ is the order of the group of automorphisms
of the polygon.)

\begin{Rem}
Observe that setting $\sfB=0$ kills $\sigma$. On the other hand, the partition
function of $\sSigma|_{\sfB=0}$ is just the torsion of the connection
$\dd_{\ev^*\sfA}$ \cite{SchwBF}. 
As a consequence, $\exp\tau(\sfa)$ is 
the perturbative expression of the torsion for
$\sfA=A_0+\sqrt\hbar\,\sfa$,
where $A_0$ is the trivial connection.
\end{Rem}

\subsection{The gauge fixing}\label{ssec-gf}
To compute $\sigma$ and $\tau$ explicitly, one has to choose a gauge fixing.
Our choice is the so-called covariant gauge fixing $\dd^\star\beta=0$,
where $\dd^\star$ is defined in terms of a Riemannian metric on $\bbr^{m-2}$,
e.g., the Euclidean metric.

In the BV formalism, one needs a gauge fixing also for some of the ghosts,
and everything has to be encoded into a gauge-fixing fermion. The first
step consists in introducing antighosts and Lagrange multipliers  
and to extend the BV action. We will denote by 
$\bar\sigma_{i,l}$ the antighosts and by $\lambda_{i,l}$ the Lagrange 
multipliers ($i=1,\dots,m-3$, $l=1,\dots,i$), with the following properties:
\begin{itemize}
\item $\overline{\sigma}_{i,l}$ is a $\frg$\ndash valued 
form of degree $m-3-i$ and ghost number $-i+2l-2$;
\item $\lambda_{i,l}$ is a $\frg$\ndash valued form of degree $m-3-i$ 
and ghost number $-i+2l-1$.
\end{itemize} 
We then introduce the corresponding antifields $\bar\sigma_{i,l}^+$
and $\lambda_{i,l}^+$.
To the BV action $\sSigma$ we add then the piece
\[
\sum_{i=1}^{m-3}\sum_{l=1}^i(-1)^i\int_{\bbr^{m-2}}
\dbraket{\bar\sigma_{i,l}^+}{\lambda_{i,l}}.
\]
By means of the Euclidean metric on $\bbR^{m-2}$, 
we can construct the corresponding Hodge $\star$ operator, which maps linearly forms on $\bbR^{m-2}$ of degree $k$ to forms of degree $m-2-k$; moreover, we 
define the $\text{L}^2$\ndash duality 
between forms on $\bbR^{m-2}$ with values in $\Lg$ and $\Lg^*$ as follows:
\begin{equation}\label{eq-L2duality}
\braket{\eta}{\omega}_{\text{L}^2}:=\int_{\bbR^{m-2}}\dbraket{\omega}{\star\eta},
\end{equation}
where the operator $\star$ acts on the form part of $\eta$. 
Finally, we choose the gauge-fixing fermion to be
\begin{multline*}
\Psi=\dbraket{\overline{\sigma}_1}{\dd^\star \beta}_{\text{L}^2}+\sum_{i=1}^{m-4} \dbraket{\overline{\sigma}_{i+1,1}}{\dd^\star \sigma_i}_{\text{L}^2}+\\
+\sum_{i=1}^{m-4}\sum_{l=2}^i \dbraket{\overline{\sigma}_{i+1,k+2-l}}{\dd^\star \overline{\sigma}_{i,l}}_{\text{L}^2}.
\end{multline*}
Observe that this gauge fixing is independent of $\sfA$, of $\sfB$ and of
the imbedding; as a consequence it is a good gauge fixing 
(according to the terminology
introduced at the end of subsection~\ref{ssec-WsBV}).
With this choice of gauge fixing, the superpropagator is readily computed.
To avoid the singularity on the diagonal of $\bbr^{m-2}\times \bbr^{m-2}$,
we prefer to work on the (open) configuration space 
\[
C_2(\bbr^{m-2}):=\{(x,y)\in\bbr^{m-2}\ |\ x\not=y\}.
\]
If we denote by $\pi_i$, $i=1,2$, the projection from 
$C_2(\bbr^{m-2})$  onto the $i$-th component, we get
\[
\left\langle 
\pi_1^*\left(\salpha^a\right)\pi_2^*\left(\sbeta_b\right)
\right\rangle_{\text{g.f.}}:=\eta\ \delta^a_b,
\]
where $\eta$  is the pullback of the normalized, 
$SO(m-2)$\ndash invariant volume form $w_{m-3}$ on $S^{m-3}$ via the map
\[
\phi\colon\begin{array}[t]{ccc}
C_2(\bbr^{m-2}) &\to & S^{m-3}\\
(x,y) &\mapsto &\frac{y-x}{\norm{y-x}}
\end{array},
\]
where $\norm{\ }$ denotes the Euclidean norm.

\subsection{Explicit expressions}
We are now in a position to write down
$\sigma$ and $\tau$ in an explicit way.
We only need a few more pieces of notation. First, we introduce the (open) configuration space $C_n(\bbR^{m-2})$ as the space of $n$ distinct points
on $\bbR^{m-2}$:
\[
C_n(\bbR^{m-2}) := \{(x_1,\dots,x_n)\in (\bbR^{m-2})^n \ | \ 
i\not=j\Rightarrow x_i\not=x_j\}.
\]
For a given $C_n$, we introduce the projections
\[
\pi_i\colon\begin{array}[t]{ccc}
C_n(\bbR^{m-2})&\to&\bbR^{m-2}\\
(x_1,\dots,x_n)&\mapsto& x_i
\end{array},
\]
and, for $i\not=j$,
\[
\pi_{ij}\colon\begin{array}[t]{ccc}
C_n(\bbR^{m-2})&\to& C_2(\bbR^{m-2})\\
(x_1,\dots,x_n)&\mapsto& (x_i,x_j)
\end{array}.
\]
Then we set
\[
\sfa_i := \left(\ev\circ(\mathrm{id}\times\pi_i)\right)^*\sfa,\quad 
\sfB_i := \left(\ev\circ(\mathrm{id}\times\pi_i)\right)^*\sfB,
\]
and
\[
\eta_{ij} := \pi_{ij}^*\,\eta.
\]
Finally, we may write
\begin{subequations}\label{expltausigma}
\begin{align}
\tau_n(\sfa) &= 
\pi^n_*\tr\left[\ad(\sfa_1)\eta_{12}\ad(\sfa_2)\eta_{23}\cdots\eta_{n-1,n}\ad(\sfa_n)\eta_{n1}\right],\\
\sigma_n(\sfa,\sfB;\Xi) &=\ii\,
\pi^{n+1}_* \braket{\ad^*(\sfa_1)\eta_{12}\ad^*(\sfa_2)\eta_{23}\cdots\ad^*(\sfa_n)\eta_{n,n+1}\sfB_{n+1}}{\Xi},
\end{align}
\end{subequations}
where $\pi^n_*$ denotes the integration along the fiber corresponding to
the projection 
$\pi^n\colon C_n(\bbR^{m-2})\times\Imbsig \to\Imbsig$, and 
$\tr$ is the trace in the adjoint representation.

\section{Properties of the Wilson surface for long knots}\label{sec-wilslong}
In this section we discuss the properties of the functions $\tau$
and $\sigma$ introduced in \eqref{expltausigma}.

\begin{figure}[h!]
\begin{center}
\resizebox{9.truecm}{!}{\includegraphics{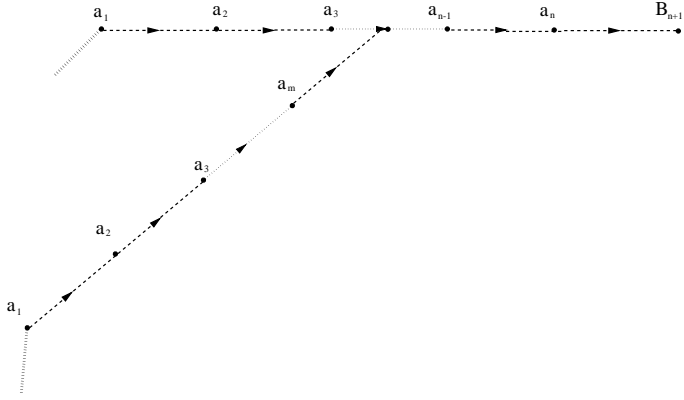}}  
\caption{The typical term in $\sbv\sigma\sigma$.}
\label{fig-sigmabra}
\end{center}
\end{figure}

\begin{figure}[h!]
\begin{center}
\resizebox{11.truecm}{!}{\includegraphics{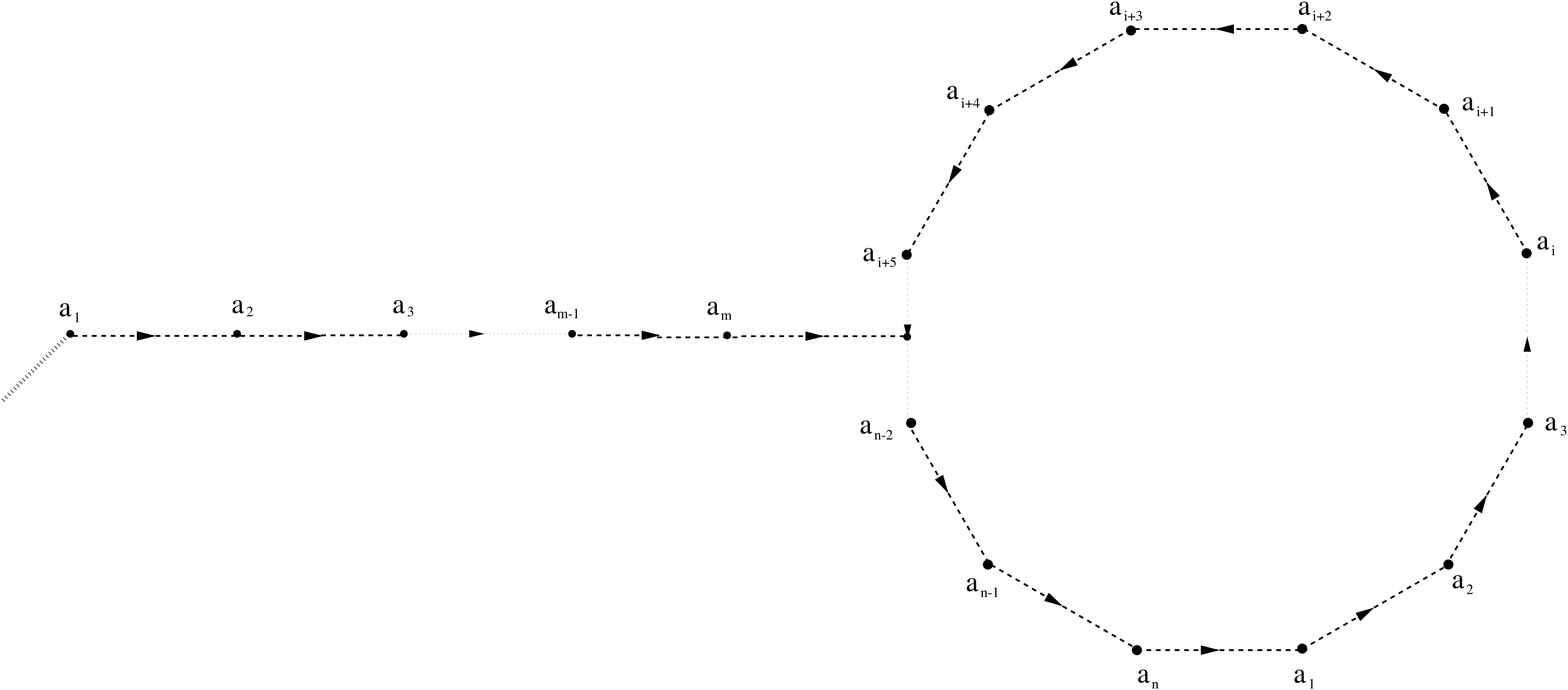}}  
\caption{The typical term in $\sfDelta\sigma$ or in $\sbv\sigma\tau$.}
\label{fig-laplbrasigmatau}
\end{center}
\end{figure}

\begin{Prop}\label{prop-tausigma}
The functions $\tau$ and $\sigma$ are well-defined and satisfy
\[
\sfdelta\tau=(-1)^m\,\dd\tau, \qquad
\sfdelta\sigma=(-1)^m\,\dd\sigma.
\]
\end{Prop}
\begin{proof}
We have first to prove that the integrals defining $\sigma$ and $\tau$
converge. This is easily done by introducing the compactifications
$C_n[\bbR^{m-2}]$ of the (open) configuration spaces $C_n(\bbR^{m-2})$
defined in \cite{BT}. These compactified configuration spaces are manifolds
with corners, with the property that all projections to
configuration spaces with less points may be lifted to smooth maps. 
Moreover, the form $\eta$ defined in the previous subsection extends to a smooth,
closed $(m-3)$\ndash form on $C_2[\bbR^{m-2}]$. As a consequence,
$\sigma$ and $\tau$ may be expressed by integrating along the compactification.
In other words, we take the same expressions but we interpret 
$\pi^n_*$ as the integration along the fiber corresponding to
the projection 
$\pi^n\colon C_n[\bbR^{m-2}]\times\Imbsig \to\Imbsig$.

To prove the properties, we use the generalized Stokes Theorem
$\dd\pi^n_*=(-1)^{mn}(\pi^n_*\dd-\pi^{n,\de}_*)$, where $\pi^{n,\de}_*$ denotes
integration along the (codimension-one) boundary of  $C_n[\bbR^{m-2}]$.
Since the forms $\eta_{ij}$ are closed, the first term produces a sum of
integrals
where $\dd$ is applied, one at a time, to a form $\sfa$ or $\sfB$.
The boundary terms may be divided into principal and hidden faces, the former
corresponding to the collapse of exactly two points. If the two points
are not consecutive, they are not joined by an $\eta$ and the integral along
the fiber vanishes by dimensional reasons. If on the other hand they are 
consecutive, the integral along the fiber of $\eta$ is normalized; we get
then contributions of the form $\lb\sfa\sfa$ or 
$\ad^*(\sfa)\sfB$. Collecting
all the terms and using \eqref{sfdelta}, 
we get the formulae displayed in the proposition, up to
hidden faces.

The vanishing of the hidden faces (corresponding to more points collapsing 
together and/or escaping to infinity) is due partly to dimensional reasons, 
partly to slight modifications of the {Kontsevich Lemma} (see~\cite{K}).
We refer the reader to \cite{Rth} for the detailed 
proof.\footnote{It should be remarked that in the proof 
we never make use of the fact that
the form $w_{m-3}$ appearing in the definition of $\eta$ (see the end of 
subsection~\ref{ssec-gf}) is $SO(m-2)$\ndash invariant; what is needed is just
that $w_{m-3}$ has the same parity of $m$ under the action 
of the antipodal map $x\mapsto-x$. Hence the proposition is still valid if,
in the definition of $\eta$, we choose $w_{m-3}$ to be any normalized
top form with the required parity under the antipodal map.}
\end{proof}
An immediate consequence of the Proposition is that the Wilson surface $\sfU$,
defined in \eqref{defsfU}, satisfies the ``semiclassical''
descent equation
\begin{equation}\label{semicldescent}
\sfdelta\sfU=(-1)^m\,\dd\sfU.
\end{equation}
In order to prove the ``quantum'' descent equation 
$\sOBV\sfU=(-1)^m\,\dd\sfU$, we must now show that, formally,
$\sfU$ is $\sfDelta$\ndash closed.
To do so, we first observe that, by the formal properties of the BV
Laplacian,
\[
\sfDelta\sfU = \sfU\,\left(
\sfDelta\sigma + \sfDelta\tau +
\frac12 \sbv\sigma\sigma+ \sbv\sigma\tau + \frac12 \sbv\tau\tau
\right).
\]
The second and last terms in parentheses vanish since $\tau$ depends only on
$\sfa$ (and not on $\sfB$). In \cite{Rth}, it is proved that also
the third term vanishes and that $\sfDelta\sigma +  \sbv\sigma\tau =0$.
Graphically, these terms are represented in fig.~\ref{fig-sigmabra} 
and~\ref{fig-laplbrasigmatau}.\footnote{The Y-shaped vertex with no labels in the figures
is the result of the contraction of an $\sfa$ with a $\sfB$ determined by the BV
bracket or the BV Laplacian.}

We observe that the proof in \cite{Rth} is rather formal in the sense that,
in the computation of $\sfDelta\sigma$, it ignores the term coming
from $\sfB$ and the adjacent $\sfa$, as this term produces a tadpole.
However, if $\frg$ is unimodular, the Lie algebraic coefficient 
of this term vanishes. In the general case, one has to introduce a suitable
counterterm $\tau_1$ in the torsion to compensate for it in
$\sbv\sigma\tau$.

Our final comment is that it does not make much sense to spend efforts in
making the proof of the quantum descent equation more rigorous. In any case,
the descent equation implies only formally that $\vev{\sfU_0}$ should be an
invariant, where $\sfU_0$ denotes the piece of $\sfU$ of degree $0$ (hence, of ghost number $0$). 
What one has to do instead is to take the perturbative
expression of $\vev{\sfU_0}$ and directly  either prove  that it produces 
invariants of long knots or compute its failure (``anomaly'' 
in the language of \cite{BT})
and understand how to correct 
it. We will see examples of this in the next Section.

\section{Perturbative invariants of long higher-dimensional knots}\label{sec-pertlong}
\begin{figure}[h!]
\begin{center}
\resizebox{9.truecm}{!}{\includegraphics{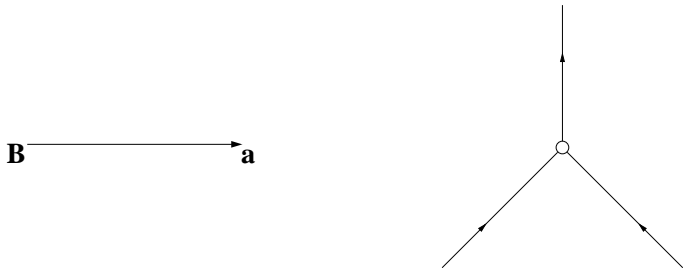}}  
\caption{The propagator and the interaction vertex.}
\label{fig-vertBF}
\end{center}
\end{figure}

In this Section we compute the first terms of the perturbative expansion of
$\vev{\sfU_0}$ and briefly discuss the expectation value of the product
of $\sfU_0$ with a Wilson loop.
First we have, however, to describe the Feynman rules for $BF$
theory. According to the action as written in \eqref{SSigma-S},
there is a superpropagator between $\sfa$ and $\sfB$,
which we will denote by a solid line oriented from $\sfB$ to
$\sfa$, and a trivalent vertex
as in fig.~\ref{fig-vertBF} of weight $\sqrt\hbar$.

In the covariant gauge, the superpropagator can easily be described as 
follows (see \cite{CR} for details).
Let us denote by $\pi_i$, $i=1,2$, the projection from 
$C_2(\bbr^{m})$  onto the $i$-th component. Then
\[
\left\langle 
\pi_1^*\left(\sfA^a\right)\pi_2^*\left(\sfB_b\right)
\right\rangle_{\text{g.f.}}:=\theta\ \delta^a_b,
\]
where $\theta$  is the pullback of the normalized, 
$SO(m)$\ndash invariant volume form $w_{m-1}$ on $S^{m-1}$ via the map
\[
\psi\colon\begin{array}[t]{ccc}
C_2(\bbr^{m}) &\to & S^{m-1}\\
(x,y) &\mapsto &\frac{y-x}{\norm{y-x}}
\end{array},
\]
with $\norm{\ }$ denoting the Euclidean norm.

In order to proceed with the discussion of the perturbative
expansion, we have to introduce some pieces of notation. Given $f\in\Imbsig$,
we denote by $C_{s,t}(f)$ the configuration space of $s+t$ points on
$\bbR^m$ the first $s$ of which are constrained to lie on the image of $f$;
in other words,
\[
C_{s,t}(f) = \left\{ 
\begin{array}{l}
(x_1,\dots,x_s)\in(\bbR^{m-2})^s\\
(y_{s+1},\dots,y_{s+t})\in(\bbR^m)^t
\end{array}
\Bigg|
\begin{array}{l}
x_i\not=x_j,\ 1\le i < j\le s\\
y_i\not=y_j,\  s <i < j\le s+t\\
f(x_i)\not=y_j,\ 1\le i\le s < j\le s+t
\end{array}
\right\}.
\]
Observe that $C_{s,0}(f)=C_s(\bbR^{m-2})$ and $C_{0,t}(f)=C_t(\bbR^m)$.
For $i,j=1,\dots,s$, $i\not=j$, we have projections
\[
\pi_{ij}\colon
\begin{array}[t]{ccc}
C_{s,t}(f) &\to & C_2(\bbR^{m-2})\\
(x_1,\dots,x_s;y_1,\dots,y_t) &\mapsto &(x_i,x_j)
\end{array}
\]
We will denote by $\eta_{ij}$ the pullback of $\eta$ by $\pi_{ij}$.
Moreover, for $i,j=1,\dots,s+t$, $i\not=j$, we have projections
\begin{equation}\label{varpi}
\varpi_{ij}\colon
\begin{array}[t]{ccc}
C_{s,t}(f) &\to & C_2(\bbR^{m})\\
(x_1,\dots,x_s;y_{s+1},\dots,y_{s+t}) &\mapsto &
\begin{cases}
(f(x_i), f(x_j)) & i,j\le s\\
(f(x_i), y_j) & i\le s < j\\
(y_i, f(x_j)) & j\le s < i\\
(y_i,y_j) & i,j > s
\end{cases}
\end{array}
\end{equation}
We will then denote  by $\theta_{ij}$ the pullback of $\theta$ by 
$\varpi_{ij}$.

As for the convergence of the integrals appearing in the perturbative 
expansion, we make the two following observations:
\begin{enumerate}
\item There are certainly divergences when a superfield $\sfa$ is paired
to a superfield $\sfB$ in the same interaction term (``tadpoles'').
The Lie algebra coefficient of tadpoles vanishes if $\frg$ is unimodular.
In general tadpoles are removed by finite renormalization.
\item The remaining terms are integrals over configuration spaces
$C_{s,t}(f)$. There exists a compactification $C_{s,t}[f]$ of these spaces
\cite{BT} such that the above projections are still smooth maps. The integrals
over the compactification then automatically converge (but do not differ
from the original ones as one has simply added a measure-zero set).
\end{enumerate}
For notational convenience in the following we will simply write $C_{s,t}$
instead of $C_{s,t}[f]$.

In the organization of the perturbative expansion, it is quite convenient to make use
of the following combinatorial
\begin{Lem}
The order in $\hbar$ equals the degree in $\Xi$.
\end{Lem}
\begin{proof}
Let us consider a Feynman diagram produced by $s_n$ snakes $\sigma_n$, $t_n$ $n$\ndash gons
$\tau_n$ and $v$ interaction vertices. We recall that $\sigma_n$ is of degree $n$ in
$\sfa$ and of degree one in $\sfB$ and in $\Xi$; $\tau_n$ is of degree $n$ in $\sfa$
and contains no $\sfB$s or $\Xi$s; each interaction vertex is of degree two in $\sfa$ and of
degree one in $\sfB$ and contains no $\Xi$.
Thus, the degree in $\Xi$ of the diagram is $\sum s_n$. Moreover,
\begin{align*}
\text{degree in }\sfa &= \sum ns_n +\sum nt_n + 2v,\\
\text{degree in }\sfB &= \sum s_n + v.
\end{align*}
By Wick's theorem these degrees must be equal, so
we get the identity
\[
\sum(n-1)s_n + \sum nt_n + v = 0.
\]
Recall now that the order in $\hbar$ of $\sigma_n$ is $(n+1)/2$, whereas the order
of $\tau_n$ is $n/2$. As the the order of each interaction vertex in $1/2$, the total
order of the diagram is
\[
\frac12\,\left(\sum(n+1)s_n + \sum nt_n + v\right)
\]
which by the previous identity is equal to $\sum s_n$. But this is also the degree
in $\Xi$.
\end{proof}

\subsection{Order $1$}
\begin{figure}[h!]
\begin{center}
\resizebox{3.truecm}{!}{\includegraphics{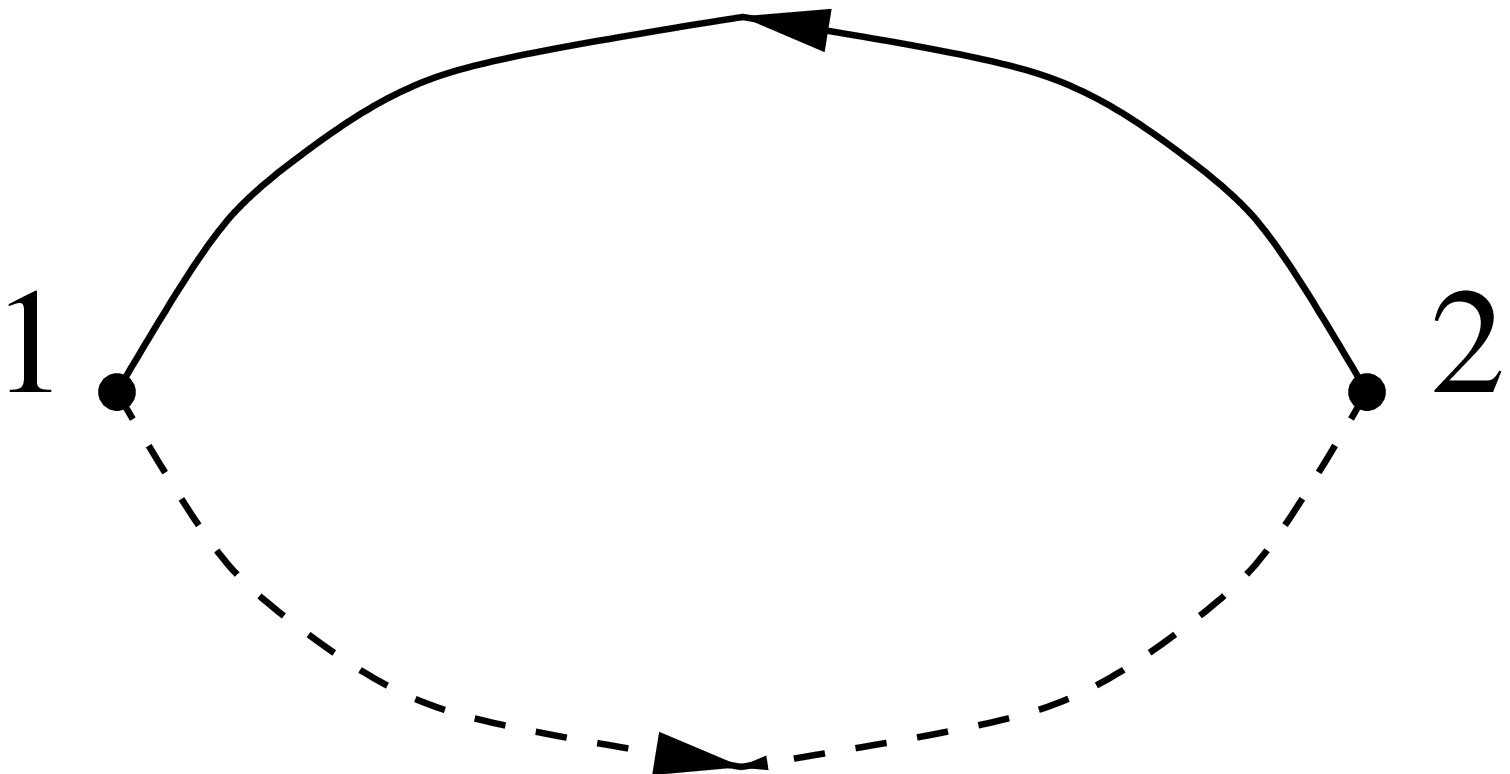}}  
\caption{Order $1$.}
\label{fig-ord1}
\end{center}
\end{figure}

The only possible term at order $1$ has the form $\Theta_1\tr(\ad\Xi)$ with
\begin{equation}\label{Theta1}
\Theta_1 = \int_{C_{2,0}} \theta_{12}\,\eta_{12}.
\end{equation}
Observe that this term does not appear
if $\frg$ is unimodular. It is also possible
to prove (considering the involution $(x_1,x_2)\mapsto(x_2,x_1)$ of
$C_{2,0}$) that $\Theta_1$ vanishes if $m$ is odd. 
The graphical representation of $\Theta_1$
is displayed in fig.~\ref{fig-ord1}. (From now on we omit in diagrams the 
black and white strip representing $\Xi$. In fig.~\ref{fig-ord1} it
would be attached to vertex $1$.)

In even dimensions, $\Theta_1$ furnishes a function on $\Imbsig$ which
is a generalization of the self-linking number for ordinary knots.
{\em This function is not an invariant}. It can be easily proved that,
in computing the differential of $\Theta_1$, the only boundary contribution
corresponds to the collapse of the two points. One obtains then
\[
\dd\Theta_1 =-p_{1*}(\Phi^*w_{m-1}\;p_3^*w_{m-3}),
\]
where
\[
\Phi\colon
\begin{array}[t]{ccc}
\Imbsig\times\bbR^{m-2}\times S^{m-3} &\to & S^{m-1}\\
(f,x,v) &\mapsto & \frac{\dd f(x)v}{\norm{\dd f(x)v}}
\end{array}
\]
and $p_i$ denotes the projection to the $i$th factor.\footnote{It may be observed
that the expression for $\dd\Theta_1$ is well-defined also when $f$
is just an immersion (and not an imbedding).  As a consequence, $\dd\Theta_1$
may be regarded as a $1$\ndash form on the space of immersions of $\bbR^{m-2}$
into $\bbR^m$ (that coincide with $\sigma$ outside a compact set).}

\subsection{Order $2$} 
\begin{figure}[h!]
\begin{center}
\resizebox{9.truecm}{!}{\includegraphics{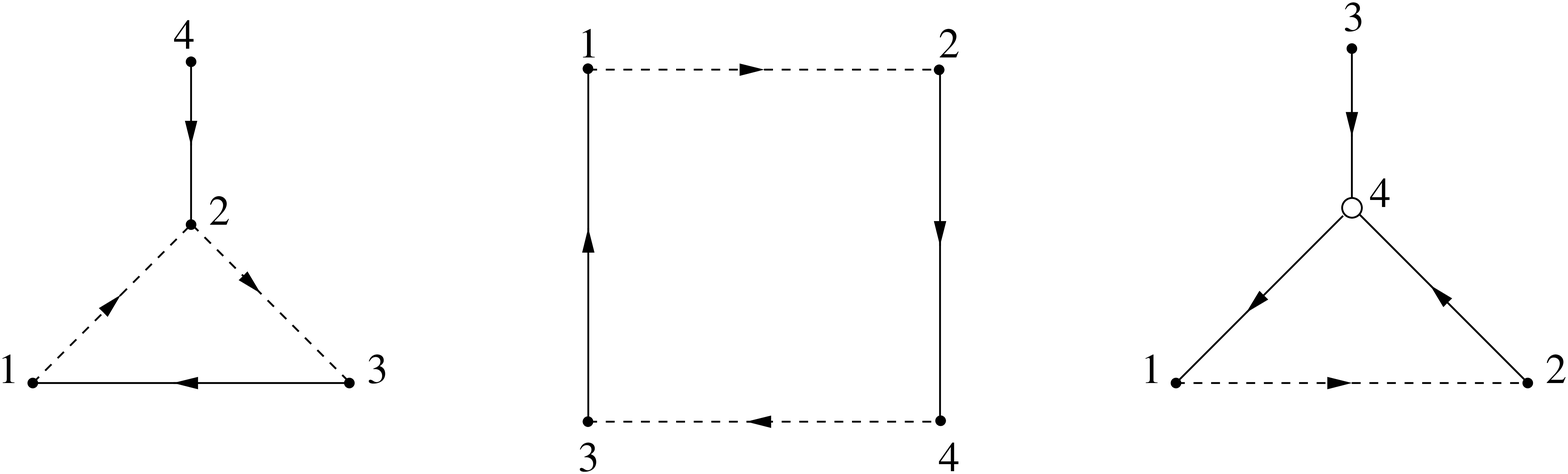}}  
\caption{Order $2$.}
\label{fig-ord2}
\end{center}
\end{figure}

The contributions corresponding to connected diagrams may be written
as $\Theta_2\tr((\ad\Xi)^2)$, where $\Theta_2$ is graphically represented in 
fig.~\ref{fig-ord2} (where white circles denote vertices in $\bbR^m$ not constrained to lie
on the image of the imbedding) 
and has the following analytical expression:
\begin{equation}\label{Theta2}
\Theta_2 = \int_{C_{4,0}} \theta_{13}\theta_{24}\eta_{12}\eta_{23} +
\frac{1}2\int_{C_{4,0}} \theta_{13}\theta_{24}\eta_{12}\eta_{34} -
\int_{C_{3,1}} \theta_{14}\theta_{24}\theta_{34}\eta_{12}.
\end{equation}
It is not difficult to prove that $\Theta_2$ vanishes if $m$ is even (consider the
involutions that exchange point $1$ with point $3$ in the first term,
point $1$ with point $2$ in the last term, and the 
pair of points $(1,3)$ with the pair $(2,4)$
in the second term).
In odd dimensions, $\Theta_2$ may be rewritten as
\[
\Theta_2=\frac{1}8\int_{C_{4,0}} \theta_{13}\theta_{24} \eta_{1234}^2-
\frac13\int_{C_{3,1}}\theta_{14}\theta_{24}\theta_{34} \eta_{123},
\]
where $\eta_{1234}$ and $\eta_{123}$ are the cyclic sums
\[
\eta_{1234}=\eta_{12}+\eta_{23}+\eta_{34}+\eta_{41},\quad \eta_{123}=\eta_{12}+\eta_{23}+\eta_{31}.
\]
In this form it is clear that  $\Theta_2$ is the long-knot version of the invariant
of knots introduced by Bott in \cite{Bott}. 
\begin{Prop}
$\Theta_2$ is an invariant.
\end{Prop}
\begin{proof}
In the computation of $\dd\Theta_2$,
the contributions of the principal faces of the three terms cancel each other as can be easily
verified. The vanishing of hidden faces may be easily proved, 
see \cite{Rth}.\footnote{Observe that also in the Chern--Simons knot invariants 
it easy to prove that hidden faces 
do not contribute to diagrams of even order.}
\end{proof}

\subsection{Order $3$} 
\begin{figure}[h!]
\begin{center}
\resizebox{13.truecm}{!}{\includegraphics{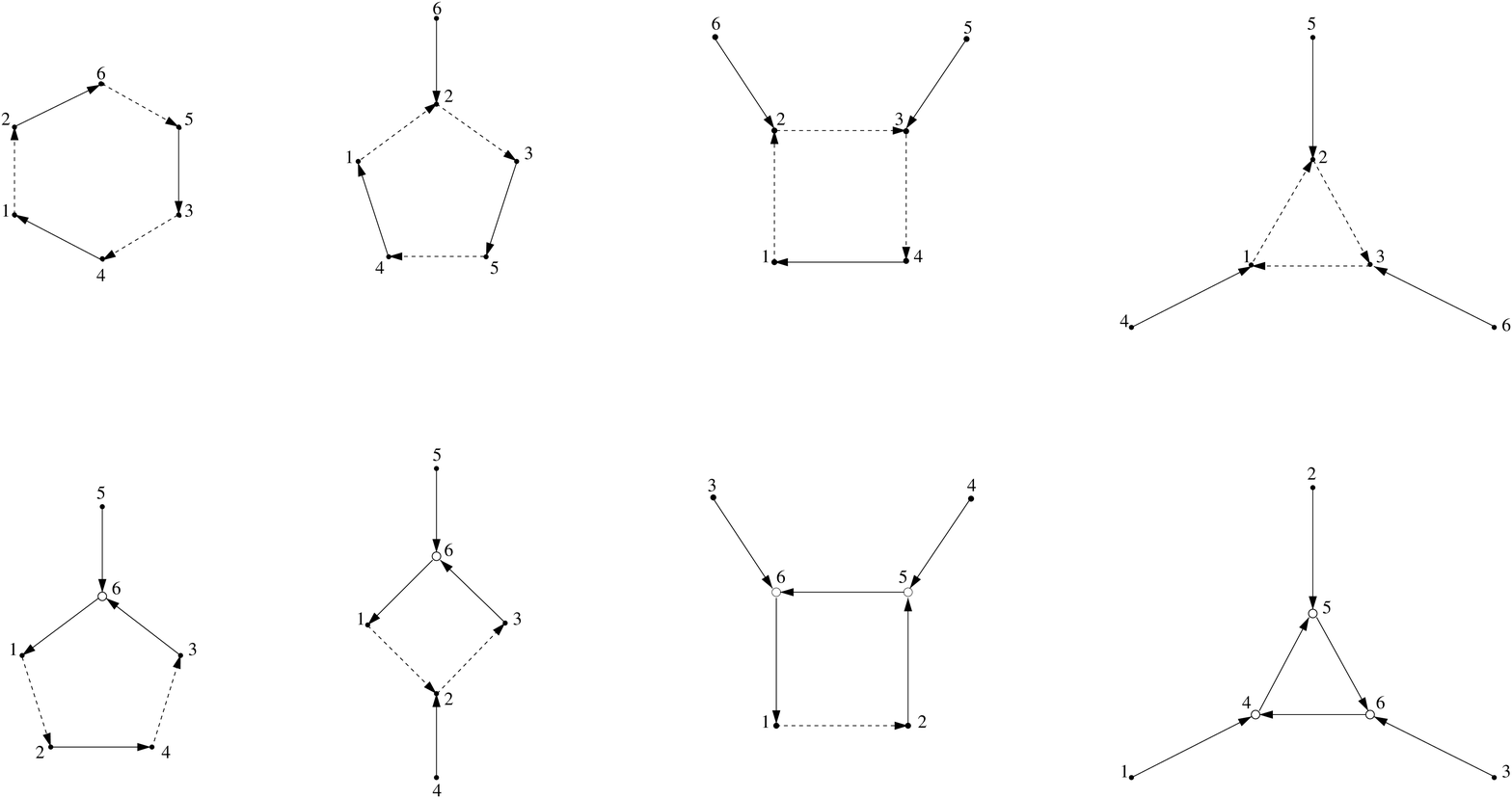}}  
\caption{Order $3$.}
\label{fig-ord3}
\end{center}
\end{figure}

Connected diagrams sum up to yield a term of the form
$\Theta_3\tr((\ad\Xi)^3)$---which clearly vanishes if the Lie algebra is 
unimodular---where $\Theta_3$ corresponds to the sum of the eight Feynman
diagrams displayed in fig.~\ref{fig-ord3}. Its analytical expression is the following:
\begin{equation}\label{Theta3}
\begin{aligned}
\Theta_3&=\frac{1}3\int_{C_{6,0}} \theta_{14}\theta_{26}\theta_{35} \eta_{12}\eta_{34}\eta_{56}+\int_{C_{6,0}} \theta_{14}\theta_{26}\theta_{35} \eta_{12}\eta_{23}\eta_{45}+\\
&\phantom{=}-\int_{C_{6,0}}\theta_{14}\theta_{26}\theta_{35} \eta_{12}\eta_{23}\eta_{34}+\frac{1}3 \int_{C_{6,0}} \theta_{14}\theta_{25}\theta_{36} \eta_{12}\eta_{23}\eta_{31}+\\
&\phantom{=}+\int_{C_{5,1}}\theta_{16}\theta_{36}\theta_{56}\theta_{24}\eta_{12}\eta_{34}
-\int_{C_{5,1}} \theta_{16}\theta_{36}\theta_{56}\theta_{24}\eta_{12}\eta_{23} +\\
&\phantom{=}-\int_{C_{4,2}}\theta_{16}\theta_{36}\theta_{56}\theta_{25}\theta_{45}\eta_{12}+\frac{1}3\int_{C_{3,3}}\theta_{14}\theta_{25}\theta_{36}\theta_{45}\theta_{46}\theta_{56}.
\end{aligned}
\end{equation}
In \cite{Rth} it is proved (by considering suitable involutions)
that $\Theta_3$ vanishes if $m$ is odd.

In even dimensions,\footnote{In this case $\Theta_3$ may also be written as
\begin{multline*}
\Theta_3=-\frac{1}{24}\int_{C_{6,0}} \theta_{14}\theta_{25}\theta_{36}\eta_{1245}\eta_{1346}\eta_{2356}-\frac{1}6\int_{C_{5,1}}\theta_{16}\theta_{36}\theta_{56}\theta_{24}\eta_{1234}\eta_{2345}+\\
-\frac{1}4\int_{C_{4,2}}\theta_{16}\theta_{36}\theta_{56}\theta_{25}\theta_{45}\eta_{1234}+\frac{1}3\int_{C_{3,3}}\theta_{14}\theta_{25}\theta_{36}\theta_{45}\theta_{46}\theta_{56},
\end{multline*}
with $\eta_{ijkl}\colon=\eta_{ij}-\eta_{jk}+\eta_{kl}-\eta_{li}$, for any 
$4$\ndash tuple of distinct indices.
In this form, $\Theta_3$ may easily be reinterpreted as a function on the space
of imbeddings of $S^{m-2}$ into $\bbR^m$.}
the differential of $\Theta_3$ is explicitly computed in \cite{Rth}
and it is proved that the only boundary
contribution that may survive in each term
is the most degenerate face, i.e., the one corresponding to the collapse
of all vertices (and with some more effort it is moreover proved that
only the seventh term may yield a nonvanishing contribution). 
To describe $\dd\Theta_3$, we first introduce the
space $\calI_{m,m-2}$ of linear injective maps from $\bbR^{m-2}$ into $\bbR^m$.
Next we consider the map
\[
T\colon
\begin{array}[t]{ccc}
\Imbsig\times\bbR^{m-2} &\to &\calI_{m,m-2}\\
(f,x) &\mapsto &\dd f(x)
\end{array}
\]
Then we may write
\[
\dd\Theta_3 = p_{1*} T^*\widehat\Theta_3,
\]
where $p_1$ is the projection onto the first factor and the ``anomaly''
$\widehat\Theta_3$ is an
$(m-1)$\ndash form that can be explicitly described as follows. Given $\alpha\in\calI_{m,m-2}$,
one defines the following action on  $C_{s,t}(\alpha)$
of the group $I=\bbR^+_*\ltimes\bbR^{m-2}$ of dilations
and translations of $\bbR^{m-2}$:
\begin{gather*}
x_i\mapsto\lambda x_i,\ i=1,\dots,s,\quad
y_i\mapsto\lambda y_i,\ i=1,\dots,t,\qquad
\lambda\in\bbR^+_*,\\
x_i\mapsto x_i+a,\ i=1,\dots,s,\quad
y_i\mapsto y_i+\alpha(a),\ i=1,\dots,t,\qquad
a\in\bbR^{m-2}.
\end{gather*}
One then defines $\widehat C_{s,t}(\alpha)$ as the quotient of $C_{s,t}(\alpha)$ by $I$.
Denoting by $\widehat C_{s,t}^{m,m-2}\to\calI_{m,m-2}$ the fiber bundle with fiber
$\widehat C_{s,t}(\alpha)$ over $\alpha$, one may write $\widehat\Theta_3$
as a sum of integrals along the fibers of $\widehat C_{s,t}^{m,m-2}$ where the integrand
form is given by the same products of propagators $\eta_{ij}$ and $\theta_{ij}$ 
as before, with the only modification that $\varpi_{ij}$, see \eqref{varpi}, 
is now defined in terms of
the linear map $\alpha$ (instead of $f$), over which the fiber lies.

In general, we do not know if $\Theta_3$ is an invariant. We briefly
describe however a possible strategy to correct it.

Let $V_{m,m-2}$ be the Stiefel manifold
(regarded as the space of linear isometries from $\bbR^{m-2}$ into $\bbR^m$ w.r.t.\ the
Euclidean metrics). Observe that $V_{m,m-2}$ is equipped with a left action
of $SO(m)$ and a free, right action of $SO(m-2)$. Let us denote by $\iota$ 
the inclusion of
$V_{m,m-2}$ into $\calI_{m,m-2}$ and by $r$ some deformation retract; viz., $r$
is a map from $\calI_{m,m-2}$ to $V_{m,m-2}$ such that $\iota\circ r$ is homotopic to the
identity (the existence of such a retract may be proved, e.g., by Gram--Schmidt
orthogonalization procedure). 
Let $h$ be a given homotopy, i.e., a map $[0,1]\times\calI_{m,m-2}\to\calI_{m,m-2}$
such that $h(0,\alpha)=\alpha$ and $h(1,\alpha)=\iota(r(\alpha))$. Define
\[
\widetilde\Theta_3 = \mathit{pr}_{2*}h^*\widehat\Theta_3,
\]
where $\mathit{pr}_{2}$ denotes the projection onto the second factor. 
Given the explicit form of $\widehat\Theta_3$, one can prove that it
is closed. Thus, we obtain
\[
\dd\widetilde\Theta_3 = -\widehat\Theta_3 + r^*\iota^*\widehat\Theta_3.
\]
It is now possible to show that $\boldsymbol\Theta_3:=\iota^*\widehat\Theta_3$ is 
$SO(m-2)\times SO(m)$\ndash invariant.\footnote{Briefly, this is true since it
is possible to extend the actions of $SO(m)$ and $SO(m-2)$ on $V_{m,m-2}$ to
the whole (restricted) bundle $\iota^*\widehat C_{s,t}^{m,m-2}\to V_{m,m-2}$
in such a way that the projection as well as the maps to $S^{m-1}$ and $S^{m-3}$ used in the
definitions of the propagators are all equivariant. Recall finally that the volume
forms $w_{m-3}$ and $w_{m-1}$ are invariant.}
If $m=4$, we can moreover prove that 
$\boldsymbol\Theta_3$
is also $SO(m-2)$\ndash horizontal; hence it is the pullback of an $SO(4)$\ndash invariant
$3$\ndash form on the Grassmannian $\mathit{Gr}_{4,2}$. Since the only such form is zero,
it follows that in four dimensions  $\boldsymbol\Theta_3=0$ and we get the following
\begin{Prop}\label{barTheta3}
$\overline\Theta_3 := \Theta_3 + p_{1*} T^*\widetilde\Theta_3$ is an invariant
of long $2$\ndash knots.
\end{Prop}
As far as we know, this invariant is new.

Observe that also for $m>4$ one may define $\overline\Theta_3$. It turns out from
the previous considerations that $\dd\overline\Theta_3= p_{1*} T^*r^*\boldsymbol\Theta_3$.
Thus, though in general we cannot claim that $\overline\Theta_3$ is an invariant,
we can compute its differential in terms of an invariant form on the Stiefel manifold.
This implies that, when $\dd\overline\Theta_3$ does not vanish, we may use it to correct
the potential invariants coming from higher-orders in perturbation theory, as explained in the
next subsection.

\subsection{Higher orders}\label{ssec-ho}
Higher-order terms may be explicitly computed. In \cite{Rth} some vanishing Lemmata are
proved which imply that only the most degenerate faces (i.e., when all points collapse)
contribute to the differential of the corresponding functions on $\Imbsig$. 
One can then prove that
in odd dimensions also these contributions vanish. One then obtains genuine invariants
of long $(m-2)$\ndash knots with $m$ odd.

In even dimensions, one may repeat the considerations of the previous subsection.
In particular, in four dimensions one may construct genuine invariants of long $2$\ndash knots.
For $m>4$, this construction yields an infinite set of functions on
$\Imbsig$ whose differentials are pullbacks of $SO(m-2)\times SO(m)$\ndash invariant
$(m-1)$\ndash forms on $V_{m,m-2}$. Since the space of such forms is 
finite dimensional \cite{GHV2}, one may produce an infinite set of invariants by taking
suitable linear combinations. This is the higher-dimensional analogue of the procedure
used in \cite{BT} to kill the (possible) anomalies in the perturbative expansion of
Chern--Simons theory with covariant gauge fixing.\footnote{In this three-\hspace{0pt}dimensional 
case, the anomaly is an $SO(3)$\ndash invariant $2$\ndash form on
the Stiefel manifold $V_{3,1}$, which may be identified with the $2$\ndash sphere. Since the space
of such forms is $1$\ndash dimensional, a single potential invariant---e.g.,
the self-linking number---is enough to correct all others.}

\subsection{Other observables}
The new observable we have introduced in this paper is not the only known observable for $BF$
theories. For example, the usual Wilson loop
\[
W_\rho(\gamma)(A)=\tr(\rho(\mathrm{Hol}(A,\gamma)),
\]
where $\rho$ is a representation of the Lie group $G$ 
and $\gamma$ an imbedding of $S^1$, is an observable; more generally,
one also has the generalized Wilson loops introduced in \cite{CCR,CR}, whose expectation values
yield cohomology classes on the space of imbeddings of $S^1$.

The expectation value of the usual Wilson loop is rather trivial (the dimension of the 
representation space) since the degree in $\sfa$ cannot be matched by the degree in $\sfB$.
The mixed expectation value of $\sfU_0$ and $W_\rho$ is more interesting. 
If $\gamma$ does not intersect $f$, the product defines an observable, and
one can show that
\begin{equation}\label{mixedvev}
\vev{\sfU_0(f)\,W_\rho(\gamma)}=\vev{\sfU_0(f)}\,\tr\ee^{-\hbar\lk(f,\gamma)\,\rho_*(\Xi)},
\end{equation}
where $\rho_*$ is the induced representation of $\frg$ and $\lk(f,\gamma)$ is the
linking number between (the images of) $f$ and $\gamma$. It can be written as
\[
\lk(f,\gamma) = \int_{\bbR^{m-2}\times S^1} \varphi^*\theta_{12} 
\]
where
\[
\mapping{\varphi}
{\bbR^{m-2}\times S^1}{C_2(\bbR^m)}{(x,t)}{(f(x),\gamma(t))}
\]

The result in \eqref{mixedvev} is tantamount saying that the only connected diagram arising from
the $n$th order in $W_\rho$ expanded in powers of $a$ is the one obtained by joining each
of these $n$ $a$s to a short snake $\sigma_0$. This result is purely combinatorial after 
observing that either joining the last $\sfa$ of a snake to the $\sfB$ of a $\sigma_0$ or
joining the two $\sfa$s of an interaction vertex to the $\sfB$s of two $\sigma_0$s yields a
factor $\Lie\Xi\Xi$ which clearly vanishes.

\section{Final comments}\label{sec-comm}
In this paper we have introduced a new observable for $BF$ theories that is associated to
imbeddings of codimension two. We list here some possible follow-ups of our work.

\subsection{Yang--Mills theory}
In \cite{BF7}, Yang--Mills theory is regarded as a deformation, called $BF$YM theory, 
of $BF$ theory
with deformation parameter the coupling constant $g_\text{YM}$.
In this setting
$\calO(A,B,f)$ becomes an observable for the $BF$YM theory 
in the topological limit $g_\text{YM}\to0$. Moreover, in this limit
the expectation value of this observable times a Wilson loop is still given by
\eqref{mixedvev}. 
Thus, $\calO$ might constitute the topological limit of a 
dual {'}t~Hooft variable \cite{tHo}.

\subsection{Nonabelian gerbes} Assume $B$ to be a two form (in the context of $BF$ theories,
we assume then that we are working in four dimensions). 
In the abelian case, the observable \eqref{abeobs} defines a connection
for the gerbe defined by $B$ \cite{Seg}; in this case, it is interesting to consider
also the case when $N$ has boundary. 
A suitable extension of our observable $\calO$ to this case would then be a candidate for
a connection on a nonabelian gerbe.

\subsection{Classical Hamiltonian viewpoint} For $M$ of the form $M_0\times\bbR$,
the reduced phase space of $BF$ theory is the space
of pairs $(A,B)$,  with $A$ a flat connection on $M_0$
and $B$ a covariantly closed $(m-2)$\ndash form of the coadjoint type, modulo symmetries.
The Poisson algebra generated by generalized Wilson loops is considered in \cite{CFP} and,
in the case $G=GL_n$ it is proved to be related to the Chas--Sullivan string topology 
\cite{CS}. It would be interesting to see which new structure one may obtain by considering
the Poisson algebra generated by generalized Wilson loops and, in addition, our new 
observables.

\subsection{Cohomology classes of imbeddings of even codimension}
In Section~\ref{sec-pertlong} we have described how the perturbative expansion 
produces
(potential) invariants of long knots. The same formulae may be used to define
forms on the space of imbeddings $\Imbsig^s$
of $\bbR^{m-2s}$ into $\bbR^m$ (with fixed linear behavior $\sigma$ at 
infinity) with $s>1$; up to hidden faces, these forms are closed (they certainly are so for 
$m$ odd). This way, we produce cohomology classes on $\Imbsig^s$. 

\subsection{Graph cohomology}
Generalizing \cite{CCL},
one can define a graph cohomology for graphs with two types of vertices
(corresponding to points on the imbedding and in the ambient space) and two types of
edges (corresponding to the two types of propagators) such that the ``integration map'' that
associates to a graph the corresponding integral over configuration spaces is a chain map 
up to hidden faces.
The Feynman diagrams discussed in this paper produce then nontrivial classes in this
graph cohomology. We plan to discuss all this in details in \cite{CRnew}.


\end{document}